\documentclass[journal]{IEEEtran}
\usepackage{amsmath,amsfonts}
\usepackage[nodayofweek]{datetime}
\usepackage{esint}
\usepackage{graphicx}
\usepackage[export]{adjustbox}
\usepackage{float}
\usepackage{cite}
\usepackage{parskip}
\usepackage{textcomp}
\usepackage{gensymb}
\usepackage{algorithm}
\usepackage{algpseudocode}
\usepackage[caption=false,font=normalsize,labelfont=sf,textfont=sf]{subfig}
\usepackage{textcomp}
\usepackage{stfloats}
\usepackage{url}
\usepackage{booktabs}
\usepackage{verbatim}
\def\BibTeX{{\rm B\kern-.05em{\sc i\kern-.025em b}\kern-.08em
    T\kern-.1667em\lower.7ex\hbox{E}\kern-.125emX}}
\usepackage{balance}
\usepackage{array, multirow}

\begin{document}

\title{Data-driven discovery of mechanical models directly from MRI spectral data}

\author{D.G.J. Heesterbeek, M.H.C. van Riel, T. van Leeuwen,  C.A.T. van den Berg, A. Sbrizzi
\thanks{Manuscript received May 1st, 2024; revised XXX XX, 2024. This work was supported by the Nederlandse Organisatie voor Wetenschappelijk Onderzoek (NWO) under grant 18897. (\textit{Corresponding author: David G.J. Heesterbeek}).}
\thanks{David G.J. Heesterbeek, Max H.C. van Riel, Cornelis A.T. van den Berg and Alessandro Sbrizzi are with the Department of Radiotherapy, Computational Imaging Group for MR Therapy and Diagnostics, University Medical Center Utrecht, 3584 CX Utrecht, The Netherlands.}
\thanks{Tristan van Leeuwen is with the Department of Mathematics, Utrecht University, 3584 CD Utrecht, The Netherlands.}
\thanks{This article has supplementary downloadable material available provided by the authors.}}

\markboth{IEEE TRANSACTIONS ON COMPUTATIONAL IMAGING, VOL.X, 2024}%
{Shell \MakeLowercase{\textit{et al.}}: A Sample Article Using IEEEtran.cls for IEEE Journals}


\maketitle

\begin{abstract}
Finding interpretable biomechanical models can provide insight into the functionality of organs with regard to physiology and disease. However, identifying broadly applicable dynamical models for in vivo tissue remains challenging. In this proof of concept study we propose a reconstruction framework for data-driven discovery of dynamical models from experimentally obtained undersampled MRI spectral data. The method makes use of the previously developed spectro-dynamic framework which allows for reconstruction of displacement fields at high spatial and temporal resolution required for model identification. The proposed framework combines this method with data-driven discovery of interpretable models using Sparse Identification of Non-linear Dynamics (SINDy). The design of the reconstruction algorithm is such that a symbiotic relation between the reconstruction of the displacement fields and the model identification is created. Our method does not rely on periodicity of the motion. It is successfully validated using spectral data of a dynamic phantom gathered on a clinical MRI scanner. The dynamic phantom is programmed to perform motion adhering to 5 different (non-linear) ordinary differential equations. The proposed framework performed better than a 2-step approach where the displacement fields were first reconstructed from the undersampled data without any information on the model, followed by data-driven discovery of the model using the reconstructed displacement fields. This study serves as a first step in the direction of data-driven discovery of in vivo models.
\end{abstract}


\newcommand*{\skipnumber}[2][1]{%
  {\renewcommand*{\alglinenumber}[1]{}\State #2}%
  \addtocounter{ALG@line}{-#1}}

\section{Introduction}

\IEEEPARstart{D}{eriving} models from first principles is generally prohibitively challenging for non-linear biomechanical systems. Data-driven discovery (DDD) has recently emerged as an attractive method for identifying interpretable closed form models from spatiotemporal measurements of complex dynamic systems\cite{schmidt_distilling_2009,  daniels_automated_2015, gonzalez_learning_2019}. The seminal work by Brunton et al.\cite{brunton_discovering_2016, rudy_data-driven_2017, kaheman_sindy-pi_2020} introduces a new approach to DDD using sparse regression to determine the underlying non-linear dynamical models directly from time-series data. This method for Sparse Identification of Non-linear Dynamics (SINDy) leverages the fact that most physical systems can be described with only a few terms making the governing equations sparse in a high-dimensional non-linear function space. Although the robustness of the method has been extensively researched using simulated data with added Gaussian noise\cite{kaheman_automatic_2022, delahunt_toolkit_2022}, its effectiveness using real measurement data is less well explored \cite{prokop_biological_2024}. 

Our goal is to use SINDy to identify dynamical models directly from experimentally obtained undersampled spectral data from the MRI scanner. MRI scanners, that naturally gather data in the spectral domain ($k$-space), provide excellent soft tissue contrast without the use of harmful radiation. However, this modality generally lacks temporal resolution at reasonable spatial resolutions required for in vivo imaging. This severely complicates the dynamical analysis, restricting the motion analysis on physiological time scales\cite{nayak_real-time_2022}. 

The spectral motion model, which has been presented as part of the recently developed Spectro-dynamic MRI framework\cite{van_riel_spectro-dynamic_2022}, is leveraged to overcome these problems. This framework enables us to distill displacement fields with a high spatiotemporal resolution directly from $k$-space data. By working directly in $k$-space, the intermediate step of reconstructing time series images (the spatial domain approach) is circumvented. This allows for the use of undersampled $k$-space data, which makes real-time imaging with high temporal resolution possible. Unlike gated approaches to dynamic MRI, where the motion is binned into different states\cite{larson_self-gated_2004}, this approach does not rely on periodicity of the dynamics. Coupling the recently introduced spectral motion model with SINDy allows for data-driven discovery of non-linear models with high temporal resolution, directly from undersampled $k$-space in an iterative manner. 

Using this approach to identify reduced order dynamical models for in vivo soft tissue holds great promise. Reduced order models are often applied to describe complex biophysical processes as they significantly reduce the system complexity\cite{pfaller_automated_2022}. Future applications of this technique for in vivo tissue include the identification of reduced order models that describe cardiac motion dynamics which are essential for understanding the functionality of the cardiac system or stomach motility for understanding an important component of the digestive process. 

The contribution of this proof of concept study is that we identify the governing equations directly from experimentally obtained undersampled $k$-space data from a dynamic phantom, acquired using a clinical MRI system. This controlled environment allows us to test and thoroughly validate the method using a series of dynamics of increasing complexity. These models include forced non-linear systems and non-linear Van der Pol oscillators previously employed to describe cardiac time series data\cite{grudzinski_modeling_2004}.

The remainder of the paper is organized as follows. In section \ref{sec: Theory} the theory behind our approach is presented, followed by an outline of the methods in section \ref{sec: Methods} in which the set-up of the MRI experiments is discussed together with the numerical implementation of the inverse problem. The main results are presented in section \ref{sec: Results} and include a comparison with a more direct 2-step approach. Open problems and possible next steps are discussed in section \ref{sec: Discussion}, ending with a conclusion in section \ref{sec: Conclusion}.

\section{Theory}\label{sec: Theory}
To provide necessary background, the building blocks of our data-driven discovery methodology are reviewed. These include the SINDy framework\cite{brunton_discovering_2016}, the employed spectral motion model\cite{van_riel_spectro-dynamic_2022} and the data consistency model\cite{van_riel_time-resolved_2023}. Afterwards, the proposed system identification method from $k$-space data is introduced.

\subsection{Data-driven discovery}\label{subsec: DDD}
Data-driven discovery as introduced by Brunton et al. \cite{brunton_discovering_2016} aims to distill the governing equations directly from time-series data of a dynamic process. For mechanical problems it is sensible to start with the general form of Newton's second law of motion: 
\begin{equation}\label{eq: dynamic equation}
    \ddot{u}_j = \frac{1}{m_{obj}}F_j(u_j, \dot{u}_j, f_{j}^{\text{ext}}, t),
\end{equation}
where $u_j$ is the displacement in the $j$'th dimension, $\dot{u}_j$ and $\ddot{u}_j$ are the first and second order time derivatives of this displacement respectively, $m_{obj}$ is the object's mass, $F_j$ is a (possibly non-linear) unknown forcing term inherent to the system in the $j$'th dimension and $f_{j}^{\text{ext}}$ is a known mass normalized external forcing term in the $j$'th dimension. A list of the symbols used in this section is presented in table \ref{tab: Nomenclature}. The goal of DDD is to find a closed form expression for $F_j$ from noisy measurements on the displacement $u_j$ that are assumed to be known. The measured displacement data for one dimension are represented as a vector in time: 
\begin{align}
    \boldsymbol{u} &= \begin{bmatrix}
           u(t_{1}) \\
           u(t_{2}) \\
           \vdots \\
           u(t_{T})
         \end{bmatrix},
\end{align}
for $T$ measurements. Here, the subscript $j$ is left out for increased readability. The same temporal discretization is performed on the (known) mass normalized external forcing, resulting in the vector $\boldsymbol{f}^{\text{ext}}$. To find the closed form expression for $F$, a finite set of candidate basis operators is introduced that $F$ might comprise of. Using $C$ basis operators, a $T\times C$ matrix $B$ is defined. An example of a matrix with 5 candidate basis operators is:
\begin{align}\label{eq: B-matrix}
    B(\boldsymbol{u}) &= \begin{bmatrix}
            \boldsymbol{u}, & \boldsymbol{u}^2, & D_t \boldsymbol{u}, & (D_t \boldsymbol{u})^2, & \boldsymbol{u}\odot D_t \boldsymbol{u}
         \end{bmatrix},
\end{align}
where $D_t$ is the temporal first order finite difference operator and $\odot$ is used for pointwise multiplication. All exponents in the context of matrix $B(\boldsymbol{u})$ are to be understood in the pointwise sense as well. These 5 candidate basis operators are a subset of the 17 monomials used in the actual reconstruction which are described in section \ref{sec: Method_Reconstruction}. Note that matrix $B$ from equation \eqref{eq: B-matrix} can contain any non-linear operator. Using the measured displacement data $\boldsymbol{u}$, we can describe the dynamic system as follows: 
\begin{equation}\label{eq: DDD}
    D_{tt}\boldsymbol{u} = B(\boldsymbol{u}) \boldsymbol{\xi} + \boldsymbol{f}^{\text{ext}}  + \boldsymbol{\eta},
\end{equation}
which is a discretized version of \eqref{eq: dynamic equation}. Here $D_{tt}$ is the temporal second order finite difference operator, $\boldsymbol{\xi}$ the $C$-dimensional vector selecting and weighting the active basis operators and $\boldsymbol{\eta}$ a vector containing noise. Here noise is to be understood in the broader sense and could include measurement errors, model imperfections, numerical errors related to the finite difference approximations, et cetera. Assuming $\boldsymbol{\eta}$ can be modeled as zero mean Gaussian noise, sparse regression can be applied to find an expression for $\boldsymbol{\xi}$, using the measured time evaluation of the displacement $\boldsymbol{u}$: 
\begin{equation}\label{eq: DDD_minimization}
    \underset{\boldsymbol{\xi}}{\arg\min} ||D_{tt}\boldsymbol{u} - \boldsymbol{f}^{\text{ext}} - B(\boldsymbol{u})\boldsymbol{\xi}||_2^2 + \lambda_S||\boldsymbol{\xi}||_1,
\end{equation}
where $\lambda_S$ denotes the weighting of the sparsity constraint. The $L^2$-norm of the residual (first term of the minimization problem) will be denoted with $\mathcal{R}(\boldsymbol{u}, \boldsymbol{\xi}, \boldsymbol{f}^{\text{ext}}, B(\boldsymbol{u}))$. 

\subsection{Spectral motion model}\label{subsec: Motion model}
MR signals are generated by nuclear spins coherently precessing under the influence of magnetic fields. The transverse magnetization $m(\vec{r}, t): \Omega_r\times\Omega_t\rightarrow\mathbb{C}$ is typically the quantity that is imaged during an MRI scan. It depends on the spatial position $\vec{r}\in\Omega_r\subseteq\mathbb{R}^{d}$ and time $t\in\Omega_t\subseteq\mathbb{R}$, where $d$ is the spatial dimension of the problem at hand ($d=2$ for this study). In steady-state, the total amount of transverse magnetization in the field of view can be considered constant under the reasonable assumption that the receive and excitation fields are homogeneous. This conservation principle leads to the continuity equation\cite{van_riel_spectro-dynamic_2022}: 
\begin{equation}\label{eq: continuity equation}
    \frac{\partial m}{\partial t} + \nabla\cdot(m \vec{v}) = 0,
\end{equation}
where $\vec{v}(\vec{r}, t): \Omega_r\times\Omega_t\rightarrow\mathbb{R}^{d}$ is the velocity field. In the remainder of this paper the notation $\vec{\cdot}$ will be used to represent physical vector fields.

During an MRI scan, additional linear gradients are applied to provide spatial encoding. This results in measurements on the spectral information of the transverse magnetization $m(\vec{r}, t)$ that are gathered on the scanner:
\begin{equation}
    \widehat{m}(\vec{k}, t)  = \mathcal{F}(m(\cdot, t))(\vec{k}) = \int_{\mathbb{R}^{d}} m(\vec{r}, t) e^{-i2\pi \vec{k} \cdot \vec{r}} \,d\vec{r}.
\end{equation}
As the transverse magnetization $m(\vec{r}, t)$ is sampled in the spectral domain, it is more natural to write \eqref{eq: continuity equation} in the spectral domain as well by applying a Fourier transform\cite{van_riel_spectro-dynamic_2022}: 
\begin{equation}\label{eq: continuity Fourier}
    \frac{\partial \widehat{m}}{\partial t} + i2\pi\sum_{j=1}^{d} k_j(\widehat{m}*\widehat{v}_j) = 0,
\end{equation}
with $\widehat{v}_j=\mathcal{F}(v_j)$ and $*$ the convolution operator. 

As \eqref{eq: continuity equation} is derived in the Eulerian (or spatial) coordinate frame, the velocity fields in this equation are also described in this framework ($\vec{v}$ in \eqref{eq: continuity equation} is $\vec{v}_{\text{Eul}}$). However, in this work we will try to find the governing equations of the displacement fields in the Lagrangian (or material) coordinate frame ($u$ in section \ref{subsec: DDD} is $u_{\text{Lag}}$). For the description of the locally rigid motion used in this work (introduced in section \ref{sec: Methods}), the partial time derivative of the displacement in the Lagrangian framework is the Eulerian velocity:
\begin{equation}\label{eq: framework}
    \frac{\partial \vec{u}_{\text{Lag}}}{\partial t} = \vec{v}_{\text{Eul}}.
\end{equation}
For computational tractability, the Lagrangian displacement fields $\vec{u}_{\text{Lag}}(\vec{r}, t):~\Omega_r~\times~\Omega_t~\rightarrow~\mathbb{R}^{d}$ are parameterized using $P$ spatial basis functions $\vec{\phi}_p(\vec{r}): \Omega_r\rightarrow\mathbb{R}^{d}$ and corresponding generalized coordinates $q_p(t): \Omega_t\rightarrow\mathbb{R}$ such that: 
\begin{equation}\label{eq: param}
    \vec{u}_{\text{Lag}}(\vec{r}, t) = \sum_{p=1}^P\vec{\phi}_p(\vec{r})q_p(t).
\end{equation}
Using prior information on the structure of the phantom used for the experiments, the number of unknowns reduces substantially with this parameterization of the displacement field. Combining \eqref{eq: continuity Fourier}, \eqref{eq: framework} and \eqref{eq: param}, the spectral motion model is derived: 
\begin{equation}\label{eq: MotionModel}
     \frac{\partial \widehat{m}}{\partial t} + i2\pi\sum_{j=1}^{d}\sum_{p=1}^{P} k_j(\widehat{m}*\widehat{\phi}_{j;p})\frac{\partial q_p}{\partial t} = 0,
\end{equation}
with $\vec{\widehat{\phi}}_p=\mathcal{F}(\vec{\phi}_p)$. Equation \eqref{eq: MotionModel} relates the $k$-space information $\widehat{m}$ to the displacement field $\vec{u}$ expressed using the predefined spatial basis functions $\vec{\phi}_p(\vec{r})$ and the generalized coordinates $q_p(t)$.

During an MRI scan, time-resolved $k$-space information $\widehat{m}$ is sampled in a discrete manner. The $k$-space domain $\Omega_k$ is discretized to an $N=N_1\times...\times N_d$ grid, and the time domain $\Omega_t$ to $T$ time points. The values of $\widehat{m}$ at the discrete points are stored in the vector $\boldsymbol{\widehat{m}}\in\mathbb{C}^{NT}$. For dynamics on physiological timescales, it is infeasible to sample the entire vector $\boldsymbol{\widehat{m}}$ as will be discussed in section \ref{sec: Data consistency model}. The discretized generalized coordinates $q_p(t)$ belonging to the spatial basis functions $\vec{\phi}_p$ are stored in the vector $\boldsymbol{q}\in\mathbb{R}^{PT}$. We define the spectral motion model term $\mathcal{G}$ as the $L^2$-norm of the residual of the discretized version of \eqref{eq: MotionModel}:
\begin{equation}\label{eq: disc_MotionModel}
    \mathcal{G}(\boldsymbol{\widehat{m}}, \boldsymbol{q}) = ||D_t\boldsymbol{\widehat{m}} + A(\boldsymbol{\widehat{m}}, \Phi)\boldsymbol{q}||_2^2,
\end{equation}
with $A$ a linear operator dependent on the $k$-space information and $\Phi$ the variable containing all information on the spatial parameterisation ($\vec{\phi}_p$ for $p\in\{1,...,P\}$). For fully sampled $k$-space data $\widehat{m}(\Omega_k, \Omega_t)$, the model in \eqref{eq: disc_MotionModel} is usually sufficient for solving the (for our spatial parameterisation) overdetermined inverse problem of finding the displacement fields, by minimizing $\mathcal{G}(\boldsymbol{\widehat{m}}, \boldsymbol{q})$ for the parameter $\boldsymbol{q}$. However, obtaining this amount of data is infeasible for subjects undergoing motion on physiological (typically sub-second) timescales. 

\subsection{Data consistency model}\label{sec: Data consistency model}
\begin{figure}[t]
    \centering
    \includegraphics[width=\columnwidth]{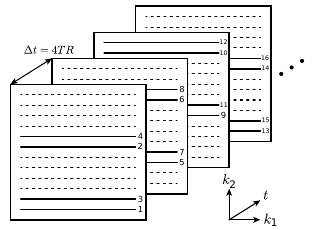}
    \caption{Schematic representation of the $k$-space sampling scheme with 12 phase encodes. The numbers indicate the temporal order in which the read-outs are sampled. The dimensions of $k$-space during the actual dynamical experiments is $|\Omega_k| = N_1\times N_2=64\times64$. In this figure $k_1$ is the read-out direction and $k_2$ the phase encode direction. For the reconstruction, 4 consecutively acquired read-out lines are grouped to form one time instance $t_i$. This results in an effective temporal resolution of $\Delta t=4\text{TR}$.}
  \label{fig: Sampling scheme}
\end{figure}

When high temporal resolution is required, hardware constraints prohibit a fully sampled read-out. To capture the motion information efficiently, a different subset of $\Omega_k$ is captured at every time instance: $\Omega_{k_{sub}}(t_i)\subseteq\Omega_k$ (see figure \ref{fig: Sampling scheme}). The acquired data contains measurement noise and is assembled in the vector $\boldsymbol{d}$. To reach sub-second time resolution, the cardinality of the set $\Omega_{k_{sub}}(t_i)$ has to be significantly lower than $N$: $|\Omega_{k_{sub}}(t_i)|=n\ll N$. To deal with the high undersampling rate and measurement noise, a data consistency model $H$ is introduced: 
\begin{equation}\label{eq: DataConsistencyModel}
    \mathcal{H}(\boldsymbol{\widehat{m}}, \boldsymbol{d}) = ||E\boldsymbol{\widehat{m}}-\boldsymbol{d}||_2^2, 
\end{equation}
with $E\in\{0, 1\}^{nT\times NT}$ a binary matrix containing the sampling pattern. 

\subsection{Proposed spectral DDD framework}
Using data-driven discovery techniques to find the governing equations that underlie the motion of objects in the MRI scanner requires measurements with a high spatiotemporal resolution. A combination of the spectral motion model \eqref{eq: disc_MotionModel}, the data consistency model \eqref{eq: DataConsistencyModel} and basic prior information about the dynamics proved to be enough to reconstruct time-resolved motion on a millisecond timescale\cite{van_riel_time-resolved_2023}. In this work we assume that no prior information about the dynamics is available, and propose a symbiosis between the SINDy framework \eqref{eq: DDD_minimization}, the spectral motion model and the data consistency model. This results in the following minimization problem: 
\begin{equation}\label{eq: minimization problem}
    \underset{\boldsymbol{\widehat{m}}, \boldsymbol{q}, \boldsymbol{\xi}}{\arg\min}~\mathcal{H}(\boldsymbol{\widehat{m}}, \boldsymbol{d}) + \lambda_{G}\mathcal{G}(\boldsymbol{\widehat{m}}, \boldsymbol{q}) + \lambda_{R}\mathcal{R}(\boldsymbol{q}, \boldsymbol{\xi}) + \lambda_S||\boldsymbol{\xi}||_1,
\end{equation}
where $\lambda_G$, $\lambda_R$ and $\lambda_S$ denote the weighting of the different terms. Note that the variables in $\mathcal{R}(\boldsymbol{u}, \boldsymbol{\xi})$ are replaced with the variables $\mathcal{R}(\boldsymbol{q}, \boldsymbol{\xi})$ after spatial parameterisation \eqref{eq: param}. The data-driven discovery term $\mathcal{R}(\boldsymbol{q}, \boldsymbol{\xi})$ can be seen as a model-based regularization on the displacement fields, that is updated every time $\boldsymbol{\xi}$ is updated. Our hypothesis is that the simultaneous interplay between these 3 models will result in a more robust and stable method for system identification compared to a more direct 2-step approach, which consists of reconstructing the displacement fields $\boldsymbol{u}$ first and subsequently performing a SINDy step. Details on the 2-step approach will be discussed in section \ref{sec: 2-step approach}, after which a comparison with the proposed method will follow in section \ref{sec: Results}. 

\begin{table}[b!]
  \centering
  \caption{Nomenclature}
  \begin{tabular}{c | c}
\toprule
\textbf{Symbol} & \textbf{Description} \\
\midrule
\multicolumn{2}{l}{\hspace{1.15cm}Notation}\\
\midrule
\textbf{bold} & Data vector\\
$\vec{\cdot}$ & Physical vector\\
$\hat{\cdot}$ & Spectral quantity\\
\midrule
\multicolumn{2}{l}{\hspace{0.8cm}Physical quantities}\\
\midrule
$t$ & Time\\
$\vec{r}/\vec{k}$ & Spatial/spectral coordinate\\
$m$ & Transverse magnetization\\
$u$ & Displacement\\
$v$ & Velocity\\
$m_{obj}$ & Object's mass\\
$\boldsymbol{\eta}$ & Stochastic noise vector\\
\midrule
\multicolumn{2}{l}{\hspace{0.9cm}Reconstruction}\\
\midrule
$B$ & Matrix containing basis operators\\
$\boldsymbol{\xi}$ & System selection vector\\
$\boldsymbol{d}$ & Acquired $k$-space data\\
$\phi$ & Spatial basis function\\
$q$ & Generalized coordinate\\
$\mathcal{R}, \mathcal{G}, \mathcal{H}$ & DDD/spectral motion/data consistency model\\
$\Omega_S, \Omega_D$ & Stationary/dynamic domain\\
\midrule
\multicolumn{2}{l}{\hspace{1cm}Operators}\\
\midrule
$D_{t(t)}$ & Temporal first (second) order finite difference operator\\
$\odot$ & Pointwise multiplication\\
$\mathcal{F}$ & Fourier transform\\
\midrule
\multicolumn{2}{l}{\hspace{1cm}Constants}\\
\midrule
$T$ & Number of measurements in time\\
$C$ & Number of basis operators\\
$P$ & Number of spatial basis functions\\
$N (n)$ & Number of (measured) $k$-space samples per time index\\
$\lambda_{G}, \lambda_{R}, \lambda_{S}$ & Weighting parameters\\
$d$ & Spatial dimension of the problem\\
\bottomrule
\end{tabular}
\label{tab: Nomenclature}
\end{table}

\section{Methods}\label{sec: Methods}
The MR-compatible Quasar\texttrademark~MRI$^{4D}$ Motion Phantom (Modus Medical Devices Inc., London, ON, Canada) was used to gather dynamical data in a controlled experimental setting. This phantom can be programmed to move a piston such that its position in time adheres to a predefined Ordinary Differential Equation (ODE). The goal is to identify the underlying ODE that governs the phantom motion with the approach presented in section 2. This approach is compared with a 2-step approach to demonstrate the superiority of the proposed method. 

\subsection{Experimental setup}\label{sec: Expermimental setup}
\begin{figure}[t]
  \centering
    \includegraphics[width=\columnwidth]{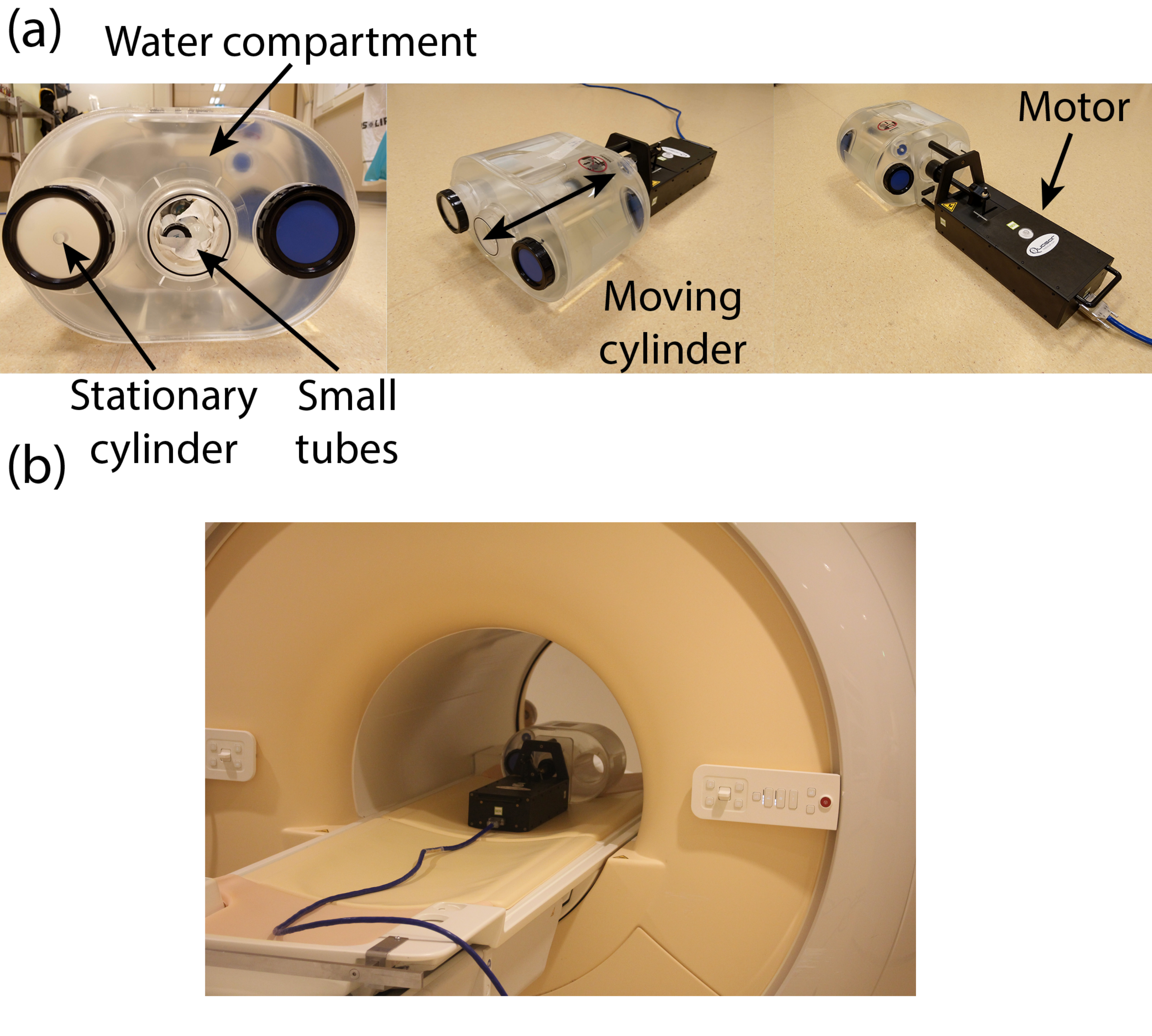}
  \caption{(a) The Quasar\texttrademark~MRI$^{4D}$ Motion Phantom. A large water compartment surrounds a stationary cylinder (off-center) and a dynamic one (center), driven by a mechanical motor. The dynamic cylinder is loaded with two smaller signal generating, gel-filled tubes.  (b) Phantom positioned in the clinical MRI scanner during the experiment.}
  \label{fig: set-up}
\end{figure}

The Quasar\texttrademark~MRI$^{4D}$ Motion Phantom is positioned in the scanner as shown in figure \ref{fig: set-up} and \ref{fig: Spatial parameterisation}. It consists of a stationary water tank with two cylindrical cavities. The central cavity acts as a guide for the cylinder driven by the mechanical motor. This moving cylinder is loaded with two gel-filled tubes (TO5, Eurospin II test system, Scotland). The off-center cavity is loaded with a stationary cylinder completely filled with gel.

The scans were performed on a 1.5T Philips Ingenia MRI scanner (Philips, Best, The Netherlands). We used a spoiled gradient echo sequence with $\text{TR}=5.5$ms, $\text{TE}=2.2$ms and a flip angle of $5\degree$. The field of view was set to be $320\times320\text{mm}^2$ in the coronal plane, with a slice thickness of 15mm. We used the body coil for data acquisition to obtain a reasonably homogeneous receive field, which is necessary for dynamical experiments. For the simple, locally rigid motion we have explored in this paper, better encoding efficiency is not required. The data sampling was performed on a $64\times64$ grid, resulting in a spatial resolution of $5\times5\text{mm}^2$. The phase encoding was played out such that two successive read-out lines were well separated in $k$-space conform the approach followed in \cite{van_riel_time-resolved_2023}. The temporal resolution of the reconstructed motion fields will be set during the reconstruction process by grouping multiple successive read-out lines to form one discrete time index $t_i$. The sampling scheme and grouping strategy are schematically visualized in figure \ref{fig: Sampling scheme} for a situation with 12 phase encodes. Grouping read-out lines results in more information per discrete time point, at the cost of a reduced temporal resolution. For our experiments 4 read-outs lines were grouped to form one discrete time point resulting in an effective temporal resolution of $\Delta t=4\text{TR}=22$ms and a $k$-space undersampling factor of $64/4=16$. The sampling pattern was cycled 40 times during each individual experiment resulting in a total of $64\times40=2560$ read-outs and an acquisition time of $14.08$s. To demonstrate that the proposed method can perform system identification along any arbitrary direction, the read-out direction ($k_1$) was rotated with $\alpha = 0\degree, 45\degree$ and $90\degree$ with respect to the direction of motion. To evaluate the repeatability, all experiments were repeated 5 times.

\subsection{Dynamics}
Motion profiles adhering to 5 different ODEs were used as position setpoint in the phantom's motion control software. These ODEs were chosen to be of increasing complexity and model order (see table \ref{tab: Dynamics}). The model order is defined as the amount of terms in the ODE excluding $\ddot{u}$ and the external forcing $f^{\text{ext}}=f(t)$. In other words, it defines the amount of active terms in the matrix $B$ (non-zero entries in the sparse $\boldsymbol{\xi}$ vector, as defined in \eqref{eq: DDD}).  The first 3 dynamics describe a forced harmonic oscillator with respectively no, linear and cubical damping. The forcing function can be found in figure S1 of the Supplementary materials and is assumed to be known. The ODEs belonging to Dynamic 4 and 5 describe different types of Van der Pol oscillators. These oscillators are used frequently in non-linear dynamics and have been shown to govern time-series data related to cardiac motion \cite{bhattacharjee_dynamics_2021, peluffo-ordonez_generalized_2015}. An improved cardiac model using the modified Van der Pol oscillator \cite{grudzinski_modeling_2004} (with the linear term left out) is used for Dynamic 5 to test for an even higher model order of 4. The specific ground truth values for the parameters can be found in table S1 of the Supplementary materials. Note the large range in the parameter values, particularly for dynamics 3-5. 

\subsection{Reconstruction}\label{sec: Method_Reconstruction}
\begin{figure}[t]
  \centering
    \includegraphics[width=\columnwidth]{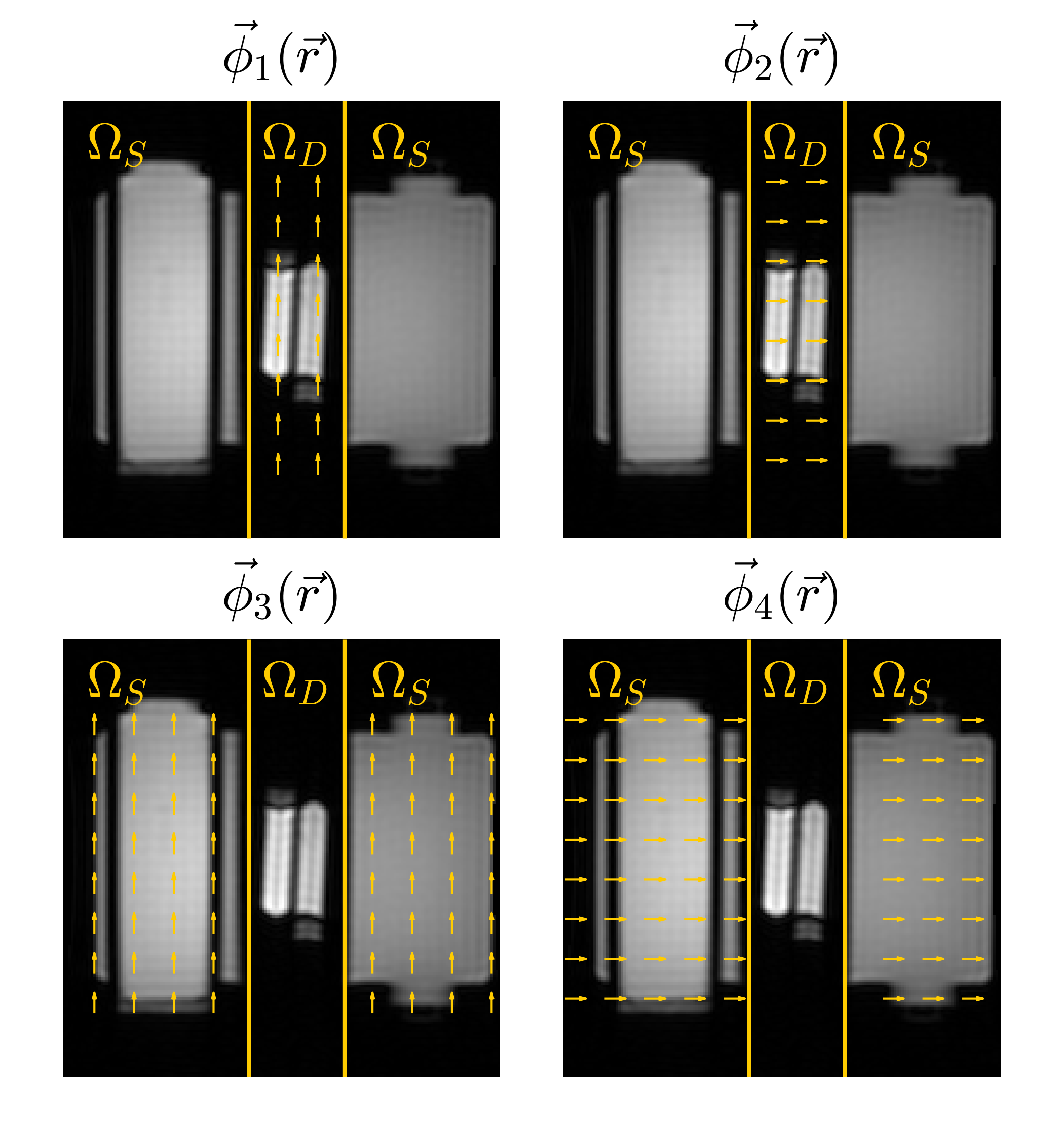}
  \caption{The 4 basis functions $\vec{\phi}(\vec{r})$ used for spatial parameterisation conform \eqref{eq: param}. Two basis functions are defined for the dynamic compartment ($\Omega_D$, top row), and two for the static compartment ($\Omega_S$, bottom row). The generalized coordinates $q_p(t)$ corresponding to each spatial basis function, are estimated during the reconstruction. Only $q_1(t)$ corresponding to basis function $\vec{\phi}_1(\vec{r})$ should be nonzero.}
  \label{fig: Spatial parameterisation}
\end{figure}

As the motion is locally rigid, the resulting motion fields can be described as spatially piecewise constant functions. Distinguishing between the stationary domain $\Omega_S$ and the dynamic domain $\Omega_D$ (see figure \ref{fig: Spatial parameterisation}), and taking into account the 2 spatial dimensions, 4 different spatial basis functions ($P=4$) are defined: 
\begin{equation}
    \begin{aligned}
    \vec{\phi}_1(\vec{r}) &= \begin{cases} (1,0)^T \qquad \forall \vec{r} \in \Omega_D \\ (0,0)^T \qquad \forall \vec{r} \in \Omega_S \end{cases} \\
    \vec{\phi}_2(\vec{r}) &= \begin{cases} (0,1)^T \qquad \forall \vec{r} \in \Omega_D \\ (0,0)^T \qquad \forall \vec{r} \in \Omega_S \end{cases} \\
    \vec{\phi}_3(\vec{r}) &= \begin{cases} (0,0)^T \qquad \forall \vec{r} \in \Omega_D \\ (1,0)^T \qquad \forall \vec{r} \in \Omega_S \end{cases} \\
    \vec{\phi}_4(\vec{r}) &= \begin{cases} (0,0)^T \qquad \forall \vec{r} \in \Omega_D \\ (0,1)^T \qquad \forall \vec{r} \in \Omega_S \end{cases}
    \end{aligned}
    \label{eq: basis-functions}
\end{equation}
Conform \eqref{eq: param}, this introduces 4 generalized coordinates $q_p(t)$ corresponding to the temporal evolution of respectively, the displacement of the dynamic compartment in the $x$ and $y$-direction and the displacement of the static compartment in the $x$ and $y$-direction. We will use $\boldsymbol{q}_{\{1,2,3,4\}}\in\mathbb{R}^T$ when referring to the discretized general coordinates corresponding to one spatial basis function (respectively $\vec{\phi}_{\{1,2,3,4\}}$).

As the cost function is block multiconvex, the minimization problem from \eqref{eq: minimization problem} can be solved for $\boldsymbol{\widehat{m}}, \boldsymbol{q}$ and $\boldsymbol{\xi}$ using the Block Coordinate Descent (BCD) method \cite{xu_block_2013}. In this iterative approach, the cost function is cyclically optimized over a subset of the variables exploiting the convexity, while the rest of the variables is kept fixed at the previously updated value. The reconstruction procedure is implemented in Matlab (The MathWorks Inc., Natick, MA, USA) and sketched in Algorithm \ref{alg: framework}. 

\begin{table}[b!]
  \centering
  \caption{Second order ordinary differential equations governing the dynamics of the motion phantom. The time dependent forcing $f(t)$ can be found in figure S1 of the Supplementary materials.}
  \begin{tabular}{c| l c c}
 Dynamic & ODE & Model order \\
\hline
1 & $\ddot{u} + a_1u = f(t)$ & 1\\
2 & $\ddot{u} + a_2u + b_2\dot{u} = f(t)$ & 2\\
3 & $\ddot{u} + a_3u + b_3\dot{u}^3 = 2f(t)$ & 2\\
4 & $\ddot{u} + a_4u + b_4\dot{u} + c_4u^2\dot{u} = 0$ & 3\\
5 & $\ddot{u} + a_5\dot{u} + b_5u^2\dot{u} + c_5u^3 + d_5u^2=0$ & 4\\
\end{tabular}
\label{tab: Dynamics}
\end{table}

The large-scale linear least-squares problem on line 4 is solved using the iterative \texttt{lsqr} solver. The non-linear least-squares problem on line 5 is solved using the \texttt{lsqnonlin} function. The non-linearity in this problem arises from the non-linear operators on $\boldsymbol{u}$ in matrix $B(\boldsymbol{u})$. Solving for $\boldsymbol{\xi}$ before pruning (line 9) is performed using an Alternating Direction Method of Multipliers (ADMM) implementation. Solving for $\boldsymbol{\xi}$ after pruning (line 12) is performed using the QR solver from the Matlab operator $\texttt{mldivide}$(\textbackslash). With the prescribed solvers, the reconstruction took approximately 15 minutes for all dynamics on a workstation with a 3.80 GHz Intel Xeon W-2235 CPU.

The (iteration dependent) hyperparameters $\lambda$, $\beta$ and $K$ are determined using a heuristic approach. For one reference dataset the hyperparameters are empirically determined by studying the algorithmic convergence. Convergence plots are presented in figure S2 of the Supplementary materials. The selected hyperparameters are subsequently fixed for all datasets and can be found in table S4 of the Supplementary materials. The main focus during hyperparameter tuning is on the residual of the data-driven learning model $\mathcal{R}$, as we consider model identification the primary aim of our reconstruction. The value of $\mathcal{R}$ indicates the agreement between the selected model ($\vec{\xi}$) and the acquired k-space data (via the reconstructed second order time derivative of the displacement ($\ddot{\vec{q}}$)).  

In the first few iterations, the estimates for $\boldsymbol{q}$ will be too noisy to achieve reasonable information on the second-order time derivative used for model identification. For this reason no estimate for $\boldsymbol{\xi}^k$ is calculated in the first $k\leq K_{\text{dyn}}$ iterations and the model-based regularization term reduces to $\mathcal{R}\left(\boldsymbol{q}, \boldsymbol{\xi}^{k-1}, b\boldsymbol{f}^{\text{ext}}, B(\boldsymbol{q})\right) = ||D_{tt}\boldsymbol{q}||_2^2$ (as $\boldsymbol{\xi}^{k-1}=\boldsymbol{0}$ and $b=0$) and acts as a regularizer for smoothness. After $K_{\text{dyn}}$ iterations, the second order derivative is assumed to have been smoothed sufficiently to start the data-driven discovery properly. As the amount of iterations increase after $k > K_{\text{dyn}}$, when DDD has commenced, the confidence in the model identification and the related model-based regularizer $\mathcal{R}(\boldsymbol{q}, \boldsymbol{\xi}^{k-1})$ increases as well. For this reason, adaptive weighting of the model-based regularizer $\mathcal{R}$ is applied on line 5 of the algorithm, using the iteration-dependent scalar $\beta^{k}$.



For model identification, we use a set of 17 monomials as candidate basis operators. These monomials are a product of powers of the displacement $u$ and the first order time derivative of the displacement $\dot{u}$. The set of candidate basis operators contains all terms required for a $4^{th}$ order Taylor expansion plus 3 additional higher order terms ($u^3\dot{u}^2$, $u^2\dot{u}^3$ and $u^3\dot{u}^3$). The matrix $B$ containing the candidate basis operators is normalized for each iteration on line 9 of the algorithm to deal with the large range of parameter values (see table S1 of the Supplementary materials) during sparse regression. This normalization is performed using the vector with $L^2$-norms: $\text{diag}\left(B(\boldsymbol{q})^{T}B(\boldsymbol{q})\right)$.

Basis operators that are not relevant for describing the dynamics are removed after $K_{\text{prune}}$ steps. This pruning of the candidate basis operators will improve the convergence of the method as the sparsity constraint on line 9 of the algorithm is known to cause underestimation of the system parameter values. The set of indices that belongs to basis operators that are pruned after $K_{\text{prune}}$ steps is denoted by $I_{\text{prune}}$. This set is created by selecting the basis operators with the lowest value for $\boldsymbol{\xi}_\text{norm}$. The amount of remaining candidate basis operators, that is $C-|I_{\text{prune}}|$, determines the model order of the discovered system and is fixed in the reconstruction.

Model identification is only performed on the generalized coordinates that reach a certain threshold after $k=K_{\text{dyn}}$ iterations to avoid model fitting when no displacement is present. For our experiments this means that $\boldsymbol{q}_1$, related to the direction of motion of the dynamic compartment of the phantom (see figure \ref{fig: Spatial parameterisation}), is reconstructed with model identification and the other 3 generalized coordinates ($\boldsymbol{q}_2$, $\boldsymbol{q}_3$ and $\boldsymbol{q}_4$) are reconstructed without model identification. For reconstruction without model identification the algorithm iteratively repeats step 3 and 4 as if $K_{\text{dyn}}>K$.

\begin{algorithm*}
\caption{Dynamic identification directly from $k$-space.}\label{alg: framework}
\begin{algorithmic}[1]
\State \textbf{Input}: $\lambda_G$, $\lambda_R$, $\lambda_S$, $\beta^k$, $K$, $K_{\text{dyn}}$, $K_{\text{prune}}$, $\boldsymbol{f}^{\text{ext}}$
\State \textbf{Initialization}: $\boldsymbol{\widehat{m}}^0\gets 0$, $\boldsymbol{q}^0\gets 0$, $\boldsymbol{\xi}^0\gets 0$, $b \gets 0$
\For{k = 1 to $K$} 
    \State $\boldsymbol{\widehat{m}}^k \gets \underset{\boldsymbol{\widehat{m}}}{\arg\min}~\lambda_{G} \mathcal{G}(\boldsymbol{\widehat{m}}, \boldsymbol{q}^{k-1}) +  \mathcal{H}(\boldsymbol{\widehat{m}}, \boldsymbol{d})$ \Comment{Reconstruction full $k$-space $\boldsymbol{\widehat{m}}$}
    \State $\boldsymbol{q}^k \gets \underset{\boldsymbol{q}}{\arg\min}~\lambda_{G} \mathcal{G}(\boldsymbol{\widehat{m}}^{k}, \boldsymbol{q}) + \lambda_{R}\beta^{k}\mathcal{R}\left(\boldsymbol{q}, \boldsymbol{\xi}^{k-1}, b\boldsymbol{f}^{\text{ext}}, B(\boldsymbol{q})\right)$ \Comment{Reconstruction displacement fields $\boldsymbol{q}$}
    \If{$k > K_{\text{dyn}}$}
    \State $b \gets 1$
    \If{$k< K_{\text{prune}}$} \Comment{Reconstruction $\boldsymbol{\xi}$ with sparsity}
        \State \mbox{$\boldsymbol{\xi}_{\text{norm}} \gets \underset{\boldsymbol{\xi}}{\arg\min}~\lambda_R \mathcal{R}\left(\boldsymbol{q}^{k}, \boldsymbol{\xi}, \boldsymbol{f}^{\text{ext}}, \text{diag}\left(B(\boldsymbol{q}^k)^{T}B(\boldsymbol{q}^k)\right)^{-1}\odot B(\boldsymbol{q}^k)\right) + \lambda_S||\boldsymbol{\xi}||_1$}
        \State $\boldsymbol{\xi}^k \gets \text{diag}\left(B(\boldsymbol{q}^k)^{T}B(\boldsymbol{q}^k)\right)^{-1}\odot \boldsymbol{\xi}_{\text{norm}}$ 
    \Else \Comment{Reconstruction $\boldsymbol{\xi}$ without sparsity using pruned set}
        \State $\boldsymbol{\xi}^k \gets\underset{\boldsymbol{\xi}}{\arg\min}~\lambda_R \mathcal{R}\left(\boldsymbol{q}^{k}, \boldsymbol{\xi}, \boldsymbol{f}^{\text{ext}}, B(\boldsymbol{q}^k)\right)$ 
        \skipnumber[1]{\phantom{$\boldsymbol{\xi}^k \gets$}} s.t. $\xi_{i\in I_{\text{prune}}} = 0$
    \EndIf
    \EndIf
\EndFor 
\end{algorithmic}
\end{algorithm*}

\subsection{Model selection}
One important aspect of data-driven discovery is the selection of the model complexity. In our work, this is represented by the model order, which is the number of active terms in the $B$ matrix (number of non-zero entries in $\boldsymbol{\xi}$). To determine the model order, we ran all reconstructions for a range of orders and plot the corresponding residuals for each of them (see figure S3 of the Supplementary materials). We select the model which shows the best balance between the obtained residual and model complexity. As seen in figure S3 of the Supplementary materials, all selected model orders are equal to the model orders of the underlying dynamics as presented in table \ref{tab: Dynamics}. 


\subsection{2-step approach}\label{sec: 2-step approach}
To benchmark the proposed joined reconstruction of the system's active terms (i.e. the sparse vector $\boldsymbol{\xi}$) and displacement $\boldsymbol{u}$, we compare the proposed method to a 2-step approach. For the 2-step approach the magnetization and displacement are reconstructed prior to the system identification step. Additional smoothness regularization is added to the displacement fields, resulting in the following 2-step optimization problem sketched below: 
\begin{align}\label{eq: two-step approach}
\begin{split}
    \text{step 1: }\hspace{1cm}&\underset{\boldsymbol{\widehat{m}}, \boldsymbol{q}}{\arg\min}~\mathcal{H}(\boldsymbol{\widehat{m}}, \boldsymbol{d}) + \lambda_{G}\mathcal{G}(\boldsymbol{\widehat{m}}, \boldsymbol{q}) + \lambda_{R}||D_{tt}\boldsymbol{q}||_2^2 \\
    \text{step 2: }\hspace{1cm}&\underset{\boldsymbol{\xi}}{\arg\min}~\lambda_{R}\mathcal{R}(\boldsymbol{q}, \boldsymbol{\xi}) + \lambda_S||\boldsymbol{\xi}||_1.
\end{split}
\end{align}
The regularization weights $\lambda$ for the 2-step approach are again determined empirically and separately from the weights of the proposed method (see table S4 of the Supplementary materials). 

\subsection{Experiments}
To validate our proposed method and show the superiority compared to the 2-step approach, the following 3 experiments were conducted. 

\subsubsection{Experiment \emph{I}}
Spectral data is gathered using our clinical MRI scanner for all 5 dynamics in table \ref{tab: Dynamics}. The proposed method as described in Algorithm \ref{alg: framework} is used to perform model identification. The corresponding estimates for the displacement are compared to the ground truth displacement programmed into the phantom. The second order time derivative of the displacement is calculated using finite difference methods. As it is directly connected to model identification via the data-driven discovery term $\mathcal{R} = ||D_{tt}\boldsymbol{q}_1 - \boldsymbol{f}^{\text{ext}} - B(\boldsymbol{q}_1)\boldsymbol{\xi}||_2^2$, it provides an additional indication of the quality of the discovered dynamics. For all experiments described in section \ref{sec: Results}, the read-out direction was parallel to the direction of motion ($\alpha=0^{\circ}$). A comparison to the acquisition setting where the read-out direction ($k_1$) was rotated with $\alpha = 45\degree$ and $90\degree$ with respect to the direction of motion is investigated in figure S4 of the Supplementary materials.

\subsubsection{Experiment \emph{II}}
Spectral data from dynamic 5 is processed using the proposed method and the 2-step approach. A comparison in terms of model identification and related second order time derivative is performed and analyzed. 

\subsubsection{Experiment \emph{III}}
To test our framework for higher undersampling factors, spectral data from dynamic 3 is retrospectively undersampled and subsequently processed using the proposed method and the 2-step approach. The retrospective undersampling is performed by cyclically keeping 20 consecutive read-outs and removing the next 40, resulting in an additional undersampling factor of 3. Combined with the already realized undersampling factor of 16 (see section \ref{sec: Expermimental setup}), this results in an effective undersampling factor of: $3\times16=48$.

\section{Results}\label{sec: Results}
The results presented in this section only relate to generalized coordinate $\boldsymbol{q}_1$. For the generalized coordinates $\boldsymbol{q}_2$, $\boldsymbol{q}_3$ and $\boldsymbol{q}_4$ the reconstructed displacements are close to $0$ as expected from the design of the phantom.

\subsection{Experiment \emph{I}}
\begin{figure}[ht!]
  \centering
    \includegraphics[width=\columnwidth]{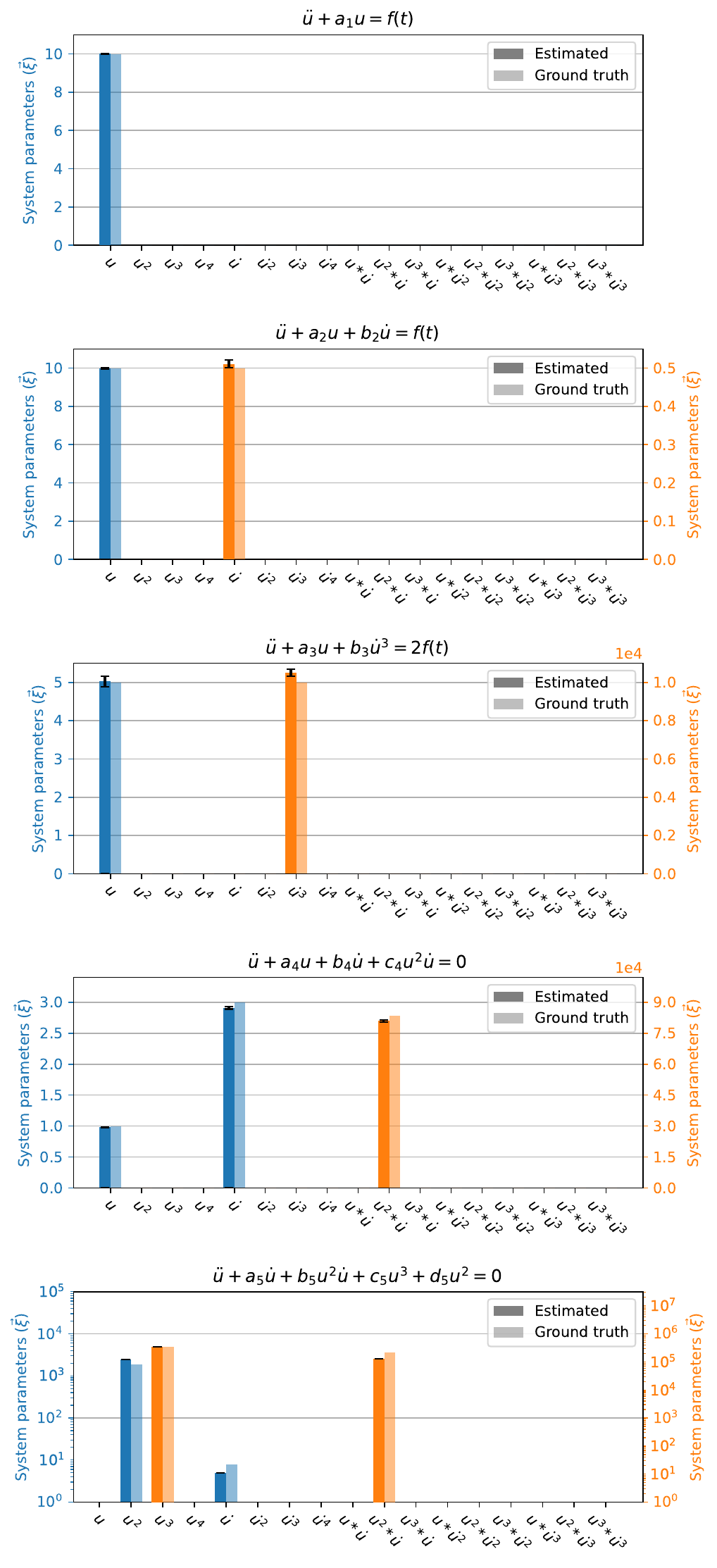}
  \caption{Model identification for dynamic 1-5. The bars refer to the absolute value of the reconstructed system parameters displayed together with the absolute value of the ground truth. The specific values for the reconstructed system parameters are reported in table S1 of the Supplementary materials. The color of the bars match the color of the corresponding $y$-axis. The error flags denote the standard deviation from 5 repetitions of the experiment.}
  \label{fig: system identification}
\end{figure}

The model identification is presented in figure \ref{fig: system identification}. In all experiments the correct basis operators have been identified with corresponding parameters for system identification close to the ground truth values. The error flags show the standard deviation from 5 repetitions of the experiment. Although the absolute value of the reconstructed system parameters is presented for readability, all estimates have been reported in table S1 of the Supplementary materials and show the correct sign conform the ground truth values. 

\begin{figure*}[t]
\centering
\begin{minipage}[t]{0.49\textwidth}
    \includegraphics[width=\columnwidth, valign=t]{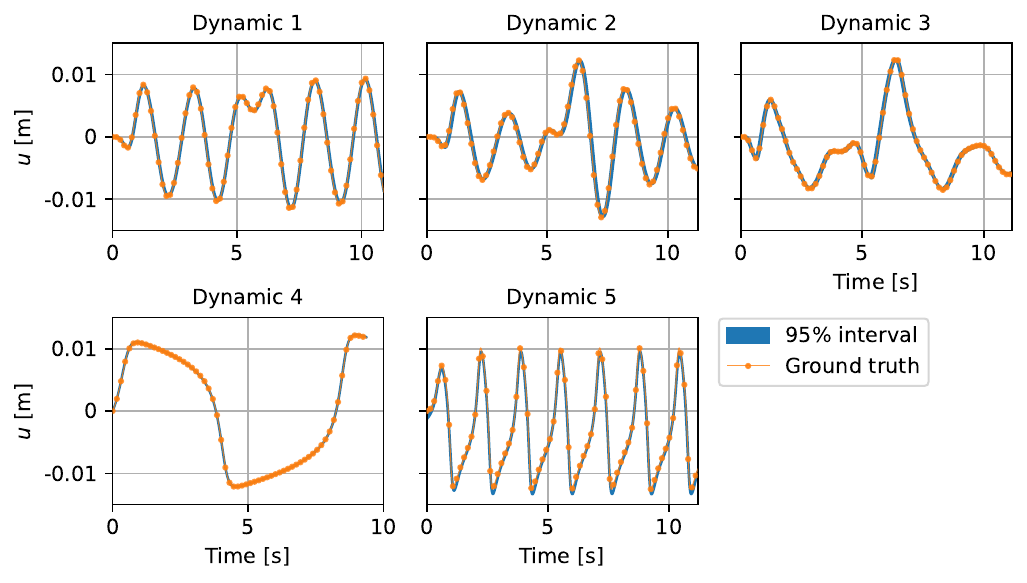}
    \caption{Estimate for the displacement ($q_1$). The 95\% interval is calculated from 5 repetitions of the experiment.}
    \label{fig: displacement}
\end{minipage}%
\hfill
\begin{minipage}[t]{0.49\textwidth}
    \includegraphics[width=\columnwidth, valign=t]{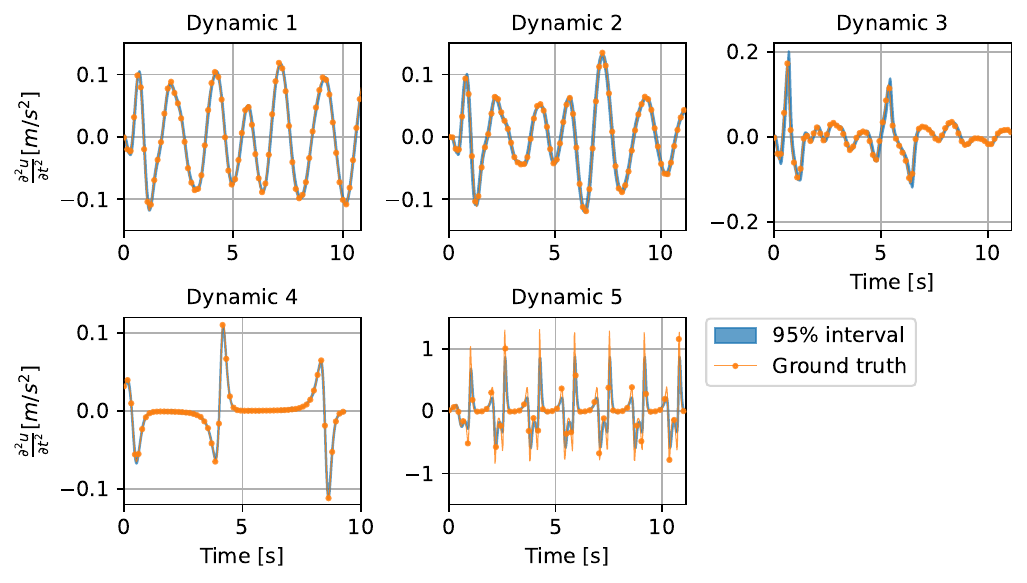}
    \caption{Estimate for the second order time derivative of the displacement ($\ddot{q}_1$). The 95\% interval is calculated from 5 repetitions of the experiment. Note the different scaling of the $y$-axis for dynamic 5 to visualize the large accelerations due to abrupt direction changes.}
    \label{fig: Second order derivative}
\end{minipage}
\end{figure*}

The estimates for the displacement ($q_1(t)$) are shown in figure \ref{fig: displacement}. Note that the 95\% interval (mean $\pm1.96\sigma$) is narrow over the entire time domain indicating high precision of the estimates. The ground truth displacement is generally close to the mean of the 95\% interval indicating that the reconstructed displacement field is accurate.

The estimates for the second order time derivative of the displacement ($\ddot{q}_1(t)$), used for model identification,  are shown in figure  \ref{fig: Second order derivative}. Again, the estimates show high accuracy and precision over the entire time domain. Only for dynamic 5 which contains very abrupt direction changes, the largest peaks get underestimated. This underestimation does not significantly deteriorate the quality of the model identification as seen in figure \ref{fig: system identification}. 

The estimates for the time-resolved images for dynamic 5 are shown in video S1 of the Supplementary materials. The time-resolved images are obtained by performing an inverse Fast Fourier Transform (IFFT) on the estimated time-resolved $k$-space data: $\mathcal{F}^{-1}(\boldsymbol{\widehat{m}})$.

Rotating the read-out direction with respect to the direction of motion with $45^{\circ}$ and $90^{\circ}$ does not significantly influence the model identification as presented in figure S4 of the Supplementary materials for dynamic 3. For all other dynamics similar results were found (not shown). 

\subsection{Experiment \emph{II}}
\begin{figure*}[b]
  \centering
    \includegraphics[width=.645\textwidth]{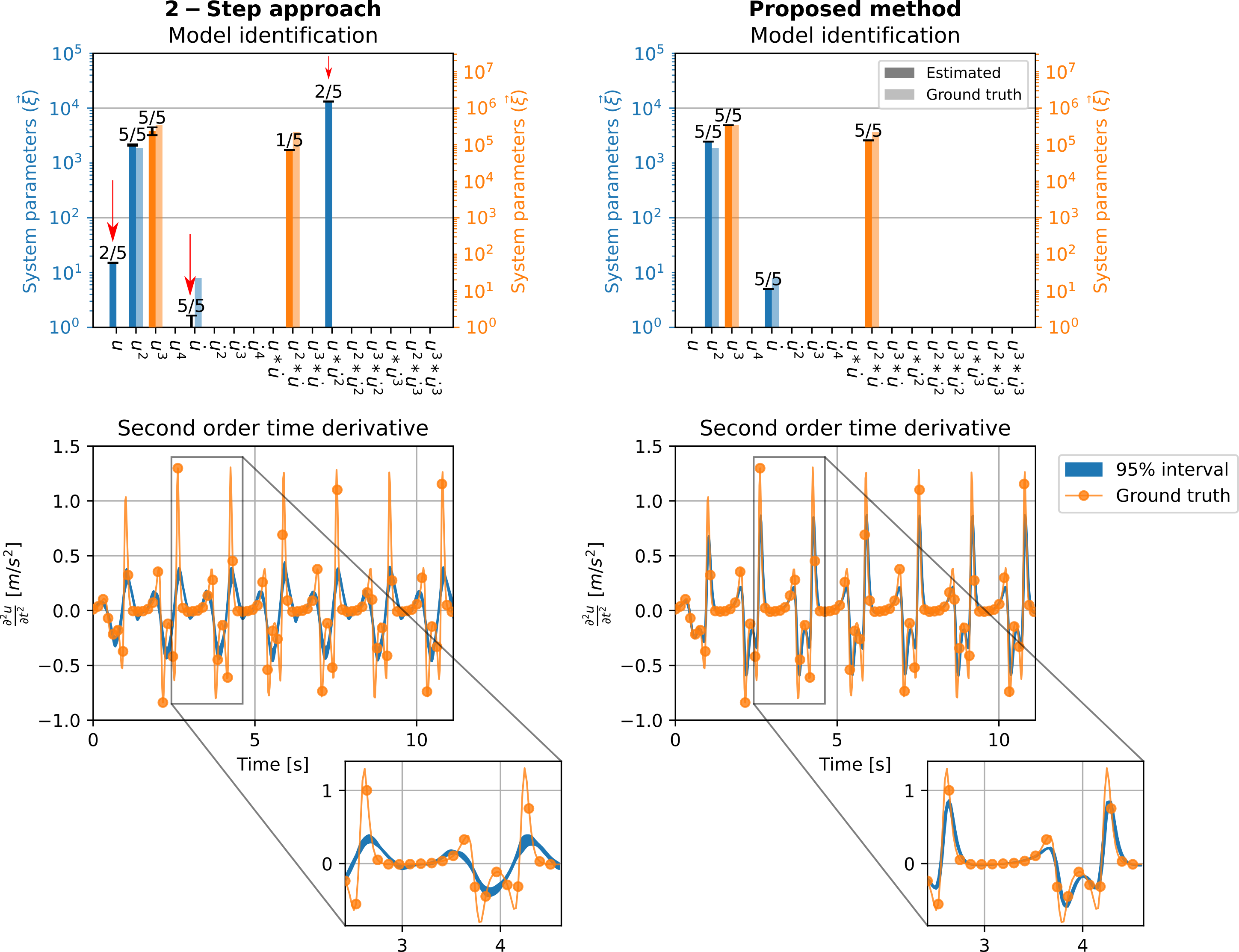}
  \caption{Model identification (displayed as in figure \ref{fig: system identification}) and related second order time derivative for dynamic 5, reconstructed using the 2-step approach and the proposed method. The specific values for the reconstructed system parameters are reported in table S2 of the Supplementary materials. The color of the bars match the color of the corresponding $y$-axis. The numbers above the bars refer to the amount of times the candidate basis operators are selected out of 5 repetitions of the experiment. The red arrows indicate the misidentified operators or operators with an unreasonable system parameter value when compared to the ground truth.}
  \label{fig: Comparison Dyn5}
\end{figure*}

In figure \ref{fig: Comparison Dyn5}, the proposed method is compared to the 2-step approach in terms of model identification and related second order time derivative for dynamic 5 at an effective temporal resolution of $\Delta t=4\text{TR}=22$ms. Note that the proposed method performs better than the 2-step approach in terms of model identification. Out of the 20 operators that have to be selected (4 operators chosen for each of the 5 repetitions) 20 are correctly identified with reasonable system parameter values (20/20) using the proposed method compared to 11/20 for the 2-step approach. Furthermore, the related estimation of the second order time derivative is significantly more accurate and precise for the proposed method. This comparison is repeated for an effective temporal resolution of $\Delta t=2\text{TR}=11$ms and the results are presented in figure S5 of the Supplementary materials. For this time resolution, the same qualitative conclusions can be drawn. 

\subsection{Experiment \emph{III}}
\begin{figure}[t]
  \centering
    \includegraphics[width=\columnwidth]{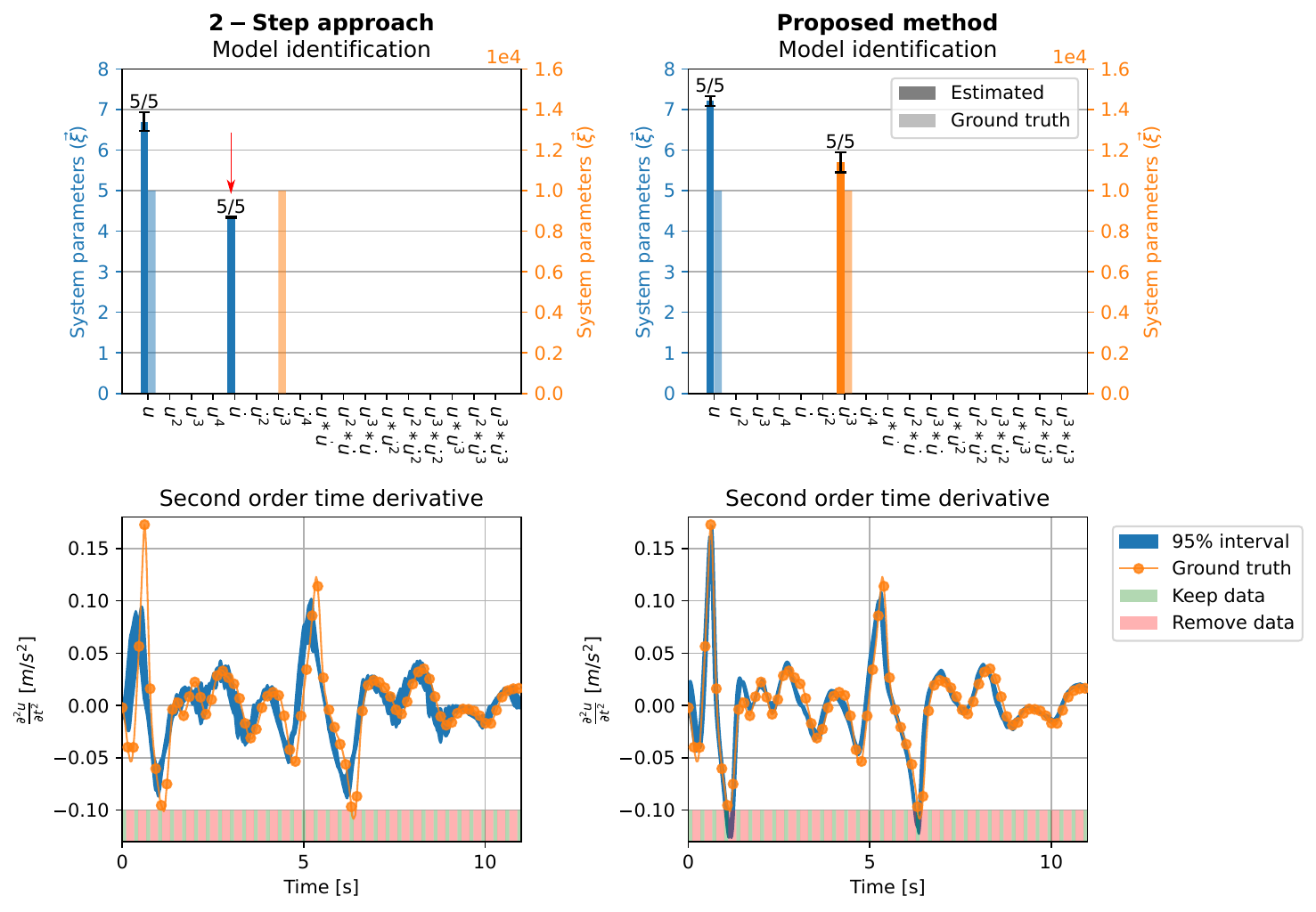}
  \caption{Model identification (displayed as in figure \ref{fig: system identification}) and related second order time derivative for dynamic 3, reconstructed using the 2-step approach and the proposed method. The specific values for the reconstructed system parameters are reported in table S3 of the Supplementary materials. The color of the bars match the color of the corresponding $y$-axis. The numbers above the bars refer to the amount of times the candidate basis operators are selected out of 5 repetitions of the experiment. Retrospective undersampling of the gathered data is performed (undersampling factor of 3). The red shading suggests the time indices for which the measured data points were removed before reconstruction. The red arrow indicates the misidentified operator.}
  \label{fig: Comparison Dyn3}
\end{figure}

In figure \ref{fig: Comparison Dyn3}, the proposed method is compared to the 2-step approach in terms of model identification and related second order time derivative for dynamic 3. In this experiment, the measured data is retrospectively undersampled in the temporal direction to add an extra undersampling factor of 3. The undersampling is sketched in figure \ref{fig: Comparison Dyn3} using red shading (remove measured spectral data at these time indices) and green shading (keep measured spectral data at these time indices). Note that the proposed method performs better compared to the 2-step approach in terms of model identification: 10/10 correctly identified compared to 5/10. Furthermore, the related estimation of the second order time derivative is significantly more accurate and precise for the proposed method. However, the reconstructed system parameter values are overestimated, especially for the operator $u$.

\section{Discussion}\label{sec: Discussion}
In this work, we proposed a reconstruction framework that simultaneously performs model identification and estimates time-resolved images and displacement fields. We formulate an optimization problem that consists of a data-driven discovery model with inherent sparsity regularization, a spectral motion model relating the displacement fields and the MRI $k$-space information and a data consistency model. This joint reconstruction problem was solved in an iterative manner, exploiting the convexity of the different sub-problems. Controlled phantom experiments in an MRI scanner showed that using our framework, accurate system identification is possible for a broad set of non-linear dynamics governed by ODEs. The method works directly from $k$-space which allows for high temporal resolution of the time-resolved images and displacement fields (22 ms in this work) that subsequently allow for system identification on this time-scale. The framework is able to deal with a broad range of system parameter values. 

Comparison with a 2-step approach, where the time-resolved images and displacement fields are reconstructed prior to the system identification step, shows the strength of the proposed, unified framework. The joined iterative reconstruction of the time-resolved images and displacement fields together with the system identification proves to be symbiotic; model information from the system identification step is leveraged for estimation of the displacement fields by using model-based regularization. This results in more accurate displacement information, which in turn benefits the system identification. For complicated quickly changing dynamics or high retrospective data undersampling factors, the differences between the methods become significant. 

We have shown that the proposed method outperforms the 2-step approach in terms of finding the correct basis operators, when considering additional undersampling. However, the applied retrospective undersampling strategy is rather aggressive as for repeated intervals of 40 time indices, no data is available. This results in an overestimation of the system parameters as shown in figure \ref{fig: Comparison Dyn3}. To deal with even higher undersampling factors, more effective strategies such as parallel imaging \cite{pruessmann_sense_1999} and compressed sensing \cite{lustig_sparse_2007} might be utilized. These strategies will become relevant for 3D implementation and we leave this analysis for future work.

One of the limitations of the method is that the convergence is dependent on a reasonable initialization for the model identification step. When the displacement fields after $K_{\text{dyn}}$ steps (see Algorithm \ref{alg: framework}) are still too far from the ground truth displacements, convergence becomes increasingly harder. Although the current empirically determined hyperparameters result in robust convergence for our datasets, more complex dynamics might require a more extensive hyperparameter optimization step \cite{akiba_optuna_2019} to assure convergence. 

Another limitation is that the time dependent external forcing of the systems is assumed to be known. For future in vivo applications this forcing, or a surrogate thereof, could be measured using e.g. an electrocardiogram (ECG) \cite{serhani_ecg_2020}, an electromyography (EMG) \cite{clarys_electromyography_1993} or a respiratory belt, depending on the application. The data-driven discovery framework is readily generalized to include external forcing by extending the operator library $B(\boldsymbol{u})$ to include the recorded actuation. This extension of the SINDy framework is described in \cite{brunton_sparse_2016}.   

An inherent limitation of SINDy is that the step of selecting the candidate basis operators is crucial to arrive at models with good predictive power. For this study, all 5 dynamics are part of the solution space spanned by the candidate basis operators. However, when the basis operators that the system consist of are not present in the operator library $B(\boldsymbol{u})$, a projection onto the function space spanned by the available basis operators can be enough for a phenomenological description if the span is sufficiently close to the model describing the dynamics. This was already explored in the Supplementary materials of the seminal paper on SINDy \cite{brunton_discovering_2016}. Furthermore, this limitation does not have to impose a major constraint on the method as the set of candidate basis operators can be arbitrarily large. 

The framework currently presented requires several extensions before it can be used for general in vivo applications. The spectral motion model derived in section \ref{subsec: Motion model} is based on the conservation of magnetization, and through-slice motion will violate this assumption. This makes an extension to 3D a logical next step when moving to in vivo applications where pure 2D displacements are rare. As the motion in vivo is generally non-rigid, the simple piecewise constant basis functions $\vec{\phi}(\vec{r})$, that are tailored to the design of the motion phantom, will have to be replaced by different spatial basis functions that describe the in vivo motion more effectively. Applying spatial parameterisation to capture deformable motion using B-splines is currently investigated by the authors since the motion fields are generally smooth almost everywhere \cite{huttinga_mr-motus_2020}. Considering deformable motion also implies changing the governing equations from ODEs to Partial Differential Equations (PDEs). The data-driven discovery framework can be extended to PDEs in a straightforward manner\cite{rudy_data-driven_2017}, such that the presented framework can remain partially unchanged.

Finally, this research can be considered as a first step in the direction of learning complete non-linear continuum mechanical descriptions \cite{flaschel_unsupervised_2021} of in vivo tissue in the form of constitutive relations. Constitutive models, relating imposed deformations to the resulting stress response of a material, form a general mathematical description of deformable mechanics. Many spatiotemporal models describing soft tissue are available in literature \cite{chagnon_hyperelastic_2015}, and using data-driven discovery combined with high quality motion fields could be of value for finding phenomenological models consistent with patient data from a broad range of subjects. These discovered models can help us to better understand soft tissue functionality with regard to physiology and disease. 


\section{Conclusion}\label{sec: Conclusion}
A reconstruction framework for model identification and estimating time-resolved displacement fields directly from spectral data has been proposed and successfully validated using experimentally obtained MRI spectral data of a dynamic phantom. The reconstructed displacement fields have a high temporal resolution required for accurate model identification without relying on periodicity of the motion. The proposed framework serves as a first step in the direction of data-driven discovery of in vivo models.


\bibliographystyle{unsrt}
\bibliography{references.bib}

\end{document}



\maketitle

\supplementarysection
\subsection{Dynamic system model and identification}
\begin{figure}[H]
  \centering
    \includegraphics[scale=0.8]{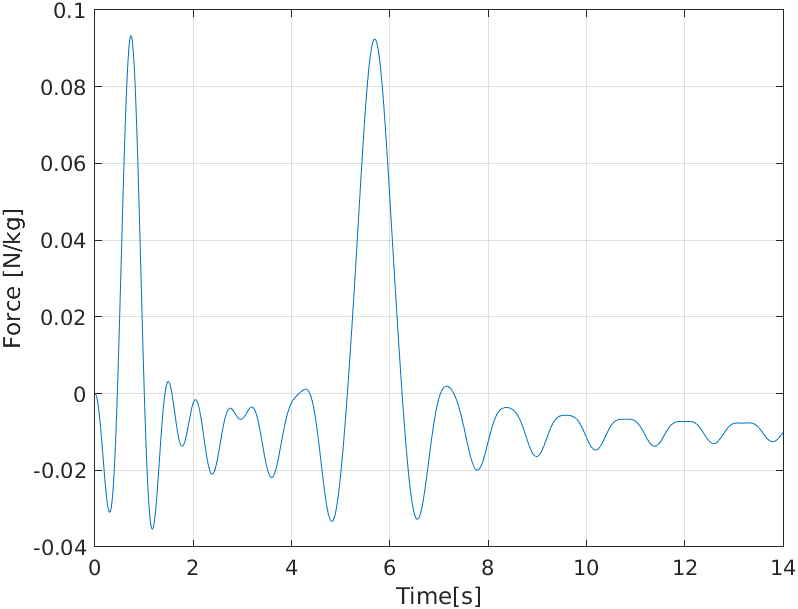}
  \caption{The external forcing $f(t)$ used in the ordinary differential equations governing the dynamics of the motion phantom. This time dependent function is generated as the sum of 2 scaled and translated sinc functions.}
  \label{supfig: forcing}
\end{figure}


\begin{table}[H]
  \centering
  \caption{Parameters used for modeling the solutions to the ordinary differential equations presented in table II of the paper, together with the reconstructed parameters and their respective standard deviations in brackets. These values correspond to the bar chart from figure 4 in the paper.}
  \scalebox{0.7}{
  \begin{tabular}{c|| l l | l l | l l | l l }
 Dynamic & $a$ & $a_{\text{Recon}}$ & $b$ & $b_{\text{Recon}}$  & $c$ & $c_{\text{Recon}}$  & $d$ & $d_{\text{Recon}}$\\
\hline
1 & 10 & $10.00(0.014)$ & $\times$ & $\times$  & $\times$ & $\times$ & $\times$ & $\times$ \\
2 & 10 & $9.98(0.032)$ & 0.5 & $0.51(0.010)$  & $\times$ & $\times$ & $\times$ & $\times$ \\
3 & 5 & $5.03(0.13)$ & $10^4$ & $1.05\cdot10^4(1.9\cdot10^2)$ & $\times$ & $\times$ & $\times$ & $\times$ \\
4 & 1 & $0.98(0.002)$ & -3 & $-2.91(0.021)$ & $8.333\cdot10^4$ & $8.09\cdot10^4(6.04\cdot10^2)$ & $\times$ & $\times$\\
5 & -8 & $-5.03(0.014)$ & $2.22\cdot10^5$ & $1.31\cdot10^5(2.04\cdot10^2)$ & $3.472\cdot10^5$ & $3.42\cdot10^5(8.84\cdot10^2)$ &$1.875\cdot10^3$ & $2.46\cdot10^3(8.05)$
\end{tabular}}
\label{suptab: ODEParameters}
\end{table}

\begin{table}[H]
  \centering
  \caption{The reconstructed parameters and their respective standard deviations in brackets for selected candidate basis functions using the 2-step approach and the proposed method. These values correspond to the bar chart from figure 7 in the paper.}
 \scalebox{0.7}{
  \begin{tabular}{c c||l l l l l l}
\multicolumn{2}{c||}{} & $u$ & $u^2$ & $u^3$ & $\dot{u}$  & $u^2*\dot{u}$ & $u*\dot{u}^2$\\
\hline
\multicolumn{2}{c||}{Ground truth value} & $\times$ & $1.875\cdot10^3$ & $3.472\cdot10^5$ & $-8$ & $2.22\cdot10^5$ & $\times$\\ 
\hline
\multirow{2}{*}{2-step approach} & mean (std) & $14.98(0.21)$ & $2.13\cdot10^3(60.7)$ & $2.40\cdot10^5(5.83\cdot10^4)$ & $-0.876(0.767)$ & $7.18\cdot10^4(0)$ & $1.32\cdot10^4(59.1)$ \\
\cline{2-2}
 & \# selected out of 5 & $2$ & $5$ & $5$ & $5$ & $1$ & $2$ \\
\hline
\multirow{2}{*}{Proposed method} & mean (std) & $\times$ & $2.46\cdot10^3(8.05)$ & $3.42\cdot10^5(8.84\cdot10^2)$ & $-5.03(0.014)$ & $1.31\cdot10^5(2.04\cdot10^2)$ & $\times$ \\
\cline{2-2}
 & \# selected out of 5 & $0$ & $5$ & $5$ & $5$ & $5$ & $0$
\end{tabular}}
\label{suptab: fig7}
\end{table}

\begin{table}[H]
  \centering
  \caption{The reconstructed parameters and their respective standard deviations in brackets for selected candidate basis functions using the 2-step approach and the proposed method. These values correspond to the bar chart from figure 8 in the paper.}
  \scalebox{0.7}{
   \begin{tabular}{c c||l l l}
\multicolumn{2}{c||}{} & $u$ & $\dot{u}$ & $\dot{u}^3$\\
\hline
\multicolumn{2}{c||}{Ground truth value} & $5$ & $\times$ & $10^4$\\
\hline
\multirow{2}{*}{2-step approach} & mean (std) & $6.69(0.21)$ & $4.33(0.018)$ & $-$ \\
\cline{2-2}
 & \# selected out of 5 & $5$ & $5$ & $0$ \\
\hline
\multirow{2}{*}{Proposed method} & mean (std) & $7.21(0.12)$ & $-$ & $1.14\cdot10^4(4.89\cdot10^2)$ \\
\cline{2-2}
 & \# selected out of 5 & $5$ & $0$ & $5$
\end{tabular}}
\label{suptab: fig8}
\end{table}

\subsection{Reconstruction parameters and convergence}
\begin{table}[H]
  \centering
  \caption{Weighting parameters used in reconstruction algorithm 1. $H$ denotes the Heaviside step function and $k$ the iteration number in the algorithm.}
  \begin{tabular}{l| l l}
 Name & Proposed method & 2-step approach\\
\hline
$K$ & 40 & 40 \\
$K_{\text{dyn}}$ & 10 & $\times$ \\
$K_{\text{prune}}$ & 35 & 35\\
$\lambda_G$ & $10^{-6}$ & $10^{-6}$ \\
$\lambda_R$ & $10^2$ & $10^2$ \\
$\lambda_S$ & $6$ & $6$ \\
$\beta^k$ & $1+9\cdot H(k-21)$ & $\times$ 
\end{tabular}
\label{suptab: Weights}
\end{table}

\begin{figure}[H]
  \centering
    \includegraphics[height=0.86\textheight]{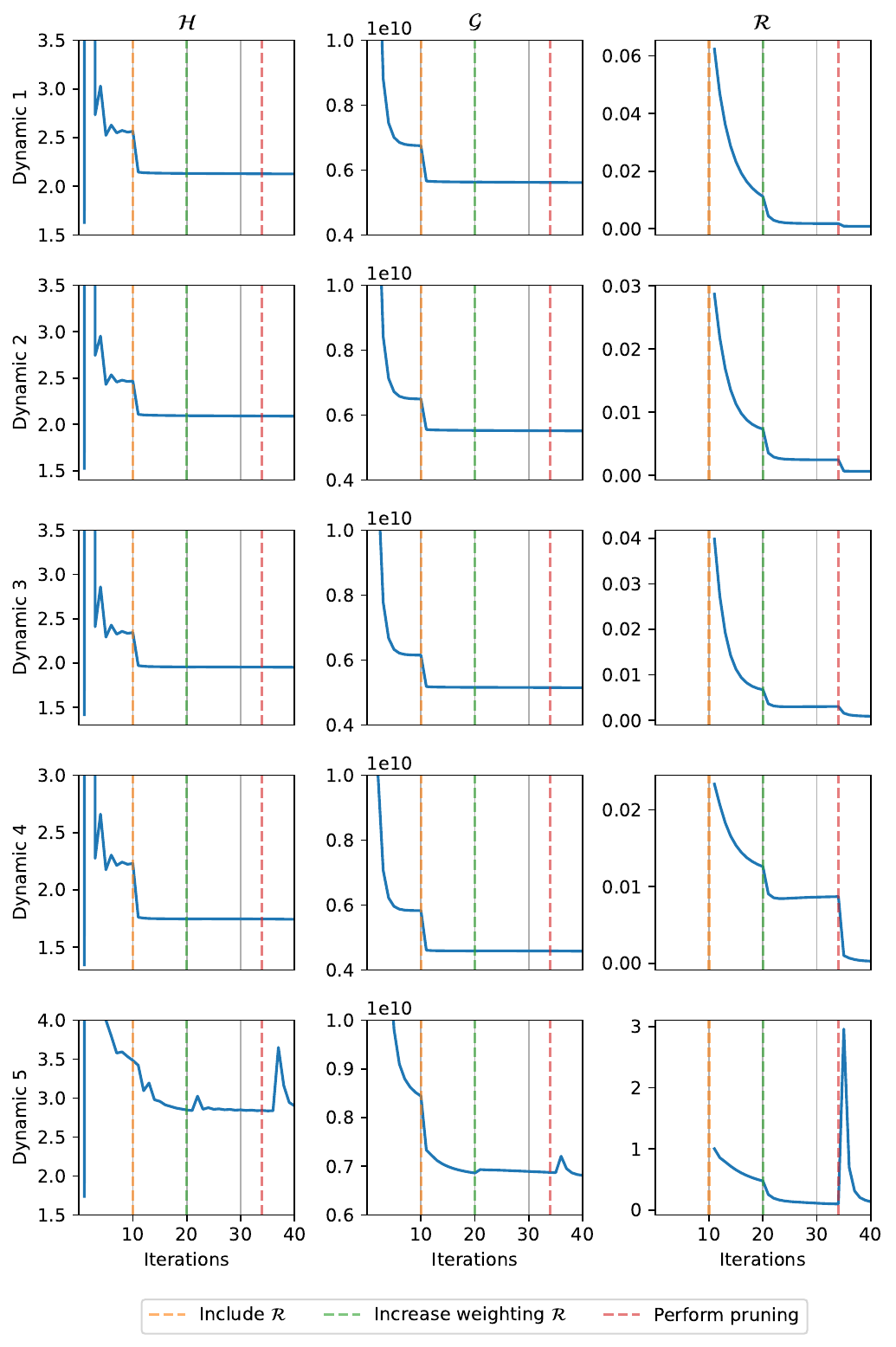}
  \caption{Convergence of the data consistency model ($\mathcal{H}$), the spectral motion model ($\mathcal{G}$) and the data-driven discovery model ($\mathcal{R}$). The vertical dashed lines denote the iterations where algorithmic changes happen that influence the convergence. The convergence for Dynamic 4 and 5 was verified by adding 10 extra iterations to the algorithm. No significant changes were observed during these extra iterations and these results are omitted from the convergence plots.}
  \label{supfig: Convergence}
\end{figure}

\subsection{Model order}

\begin{figure}[H]
\centering
\begin{subfigure}{0.45\textwidth}
  \centering
  \includegraphics[width=\linewidth]{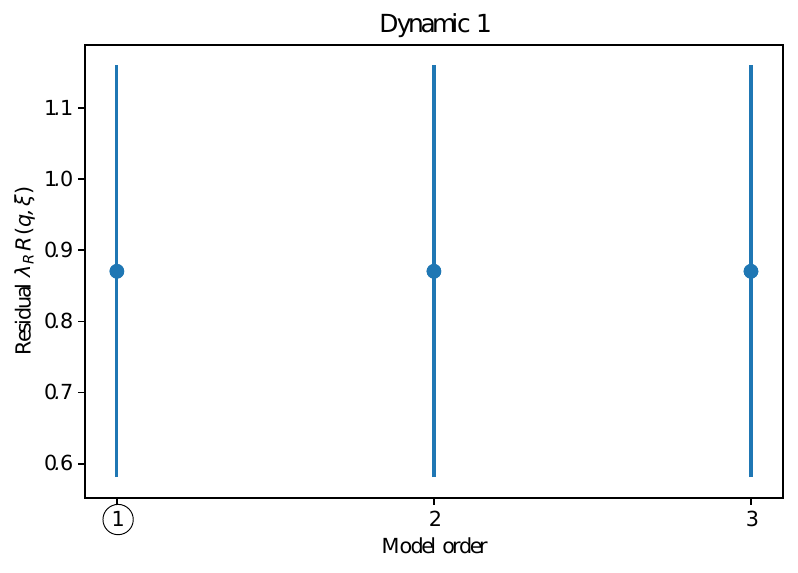}
  \label{fig:Dyn1}
\end{subfigure}%
\begin{subfigure}{0.45\textwidth}
  \centering
  \includegraphics[width=\linewidth]{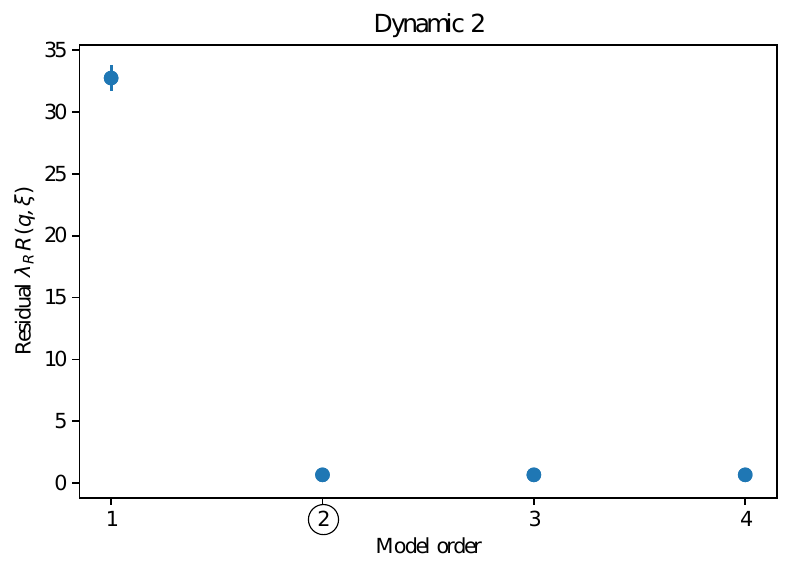}
  \label{fig:Dyn2}
\end{subfigure}
\begin{subfigure}{0.45\textwidth}
  \centering
  \includegraphics[width=\linewidth]{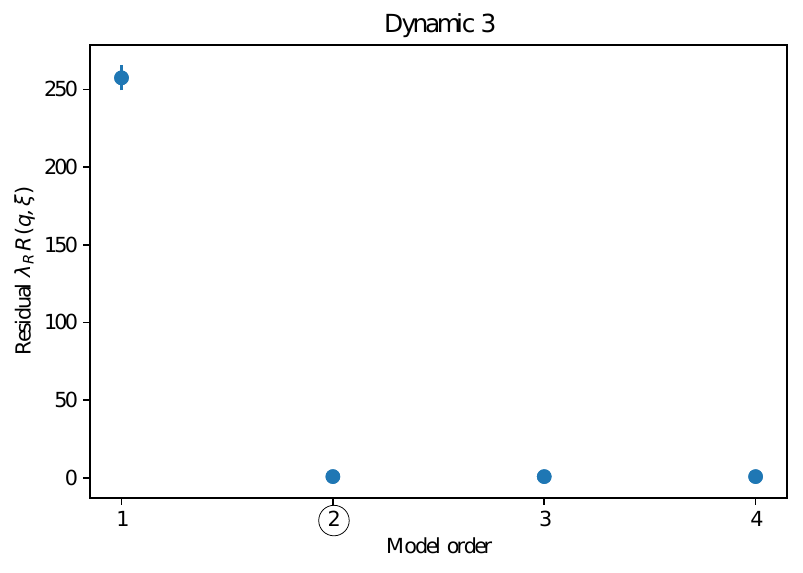}
  \label{fig:Dyn3}
\end{subfigure}%
\begin{subfigure}{0.45\textwidth}
  \centering
  \includegraphics[width=\linewidth]{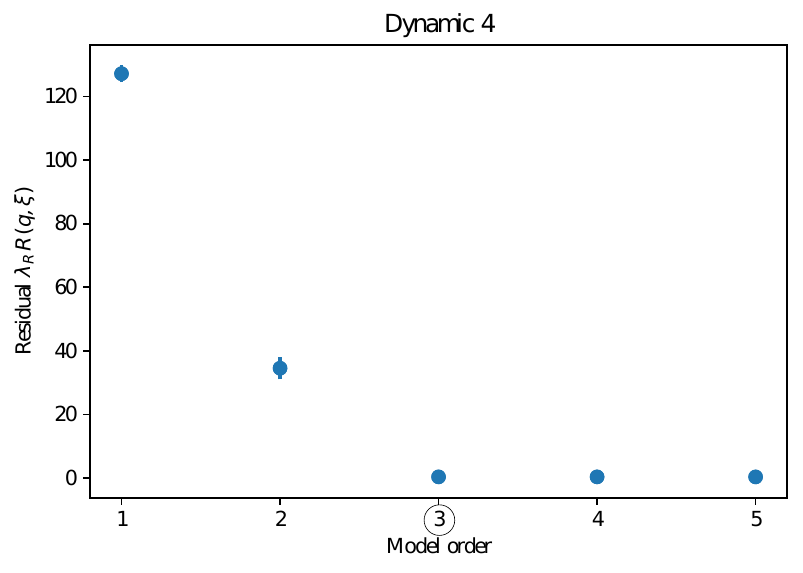}
  \label{fig:Dyn4}
\end{subfigure}%

\begin{subfigure}{0.45\textwidth}
  \centering
  \includegraphics[width=\linewidth]{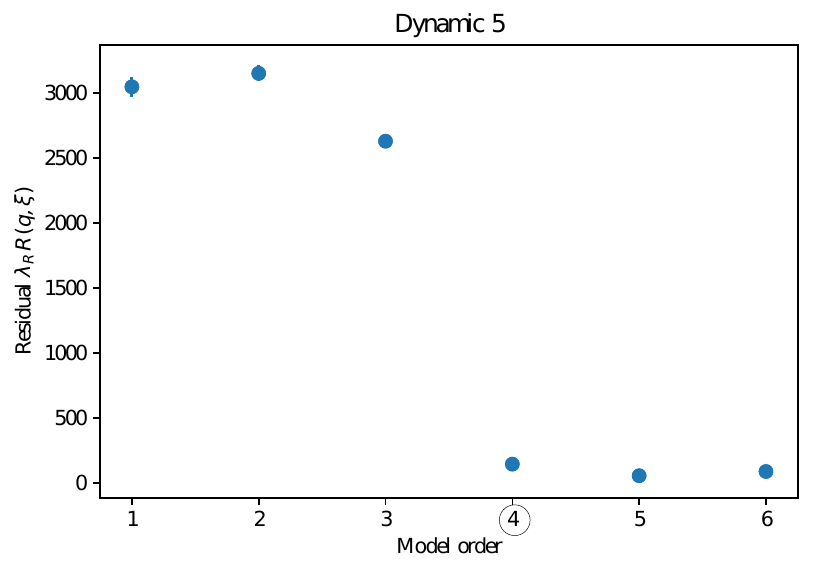}
  \label{fig:Dyn5}
\end{subfigure}%
\caption{L-curves showing the model order versus the scaled residual of the data-driven discovery model. Heuristic model order selection is performed by investigating which model order provides the best balance between predictive value and model complexity (circled). Note that the selected model orders are equal to the model orders of the underlying dynamics, presented in table II of the paper. The error flags denote the standard deviation from 5 repetitions of the experiment.}
\label{fig:modelOrder}
\end{figure}

\subsection{Read-out direction}
\begin{figure}[H]
  \centering
    \includegraphics[width=\textwidth]{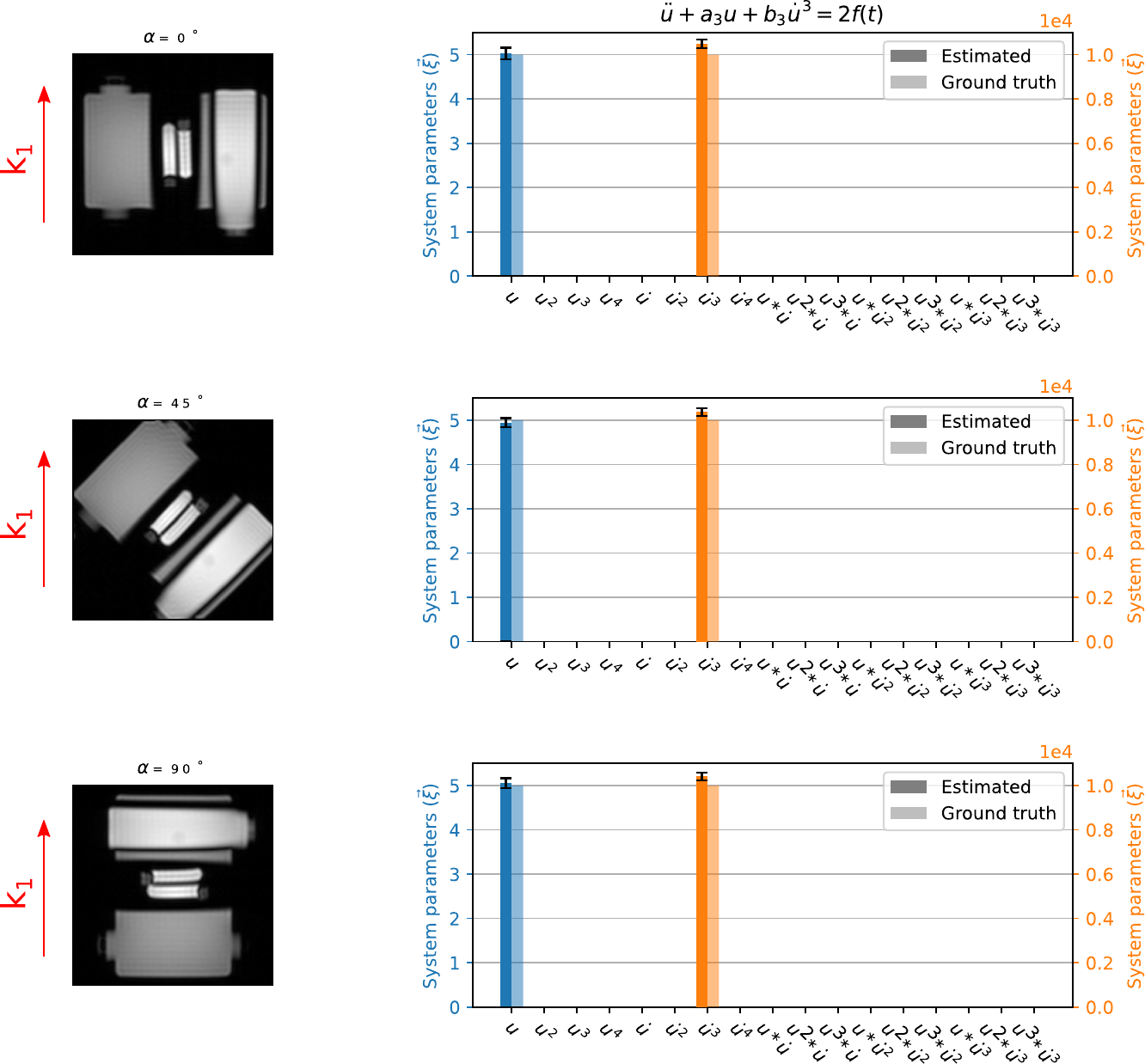}
  \caption{Model identification for dynamic 3, using 3 different read-out directions. The bars refer to the absolute value of the reconstructed system parameters together with the absolute value of the ground truth. The specific values for the reconstructed system parameters are reported in table S5. The error flags denote the standard deviation from 5 repetitions of the experiment.}
  \label{supfig: read-out direction}
\end{figure}

\begin{table}[H]
  \centering
  \caption{The reconstructed parameters and their respective standard deviations in brackets for dynamic 3. These values correspond to the bar chart from figure S4.}
  {
  \begin{tabular}{c|| l l }
  & $a_{3}$ & $b_{3}$ \\
\hline
Ground truth & $5$ & $10^4$ \\
\hline
$\alpha=0^{\circ}$ & $5.03(0.133)$ & $1.05\cdot10^4(1.94\cdot10^2)$ \\
$\alpha=45^{\circ}$ & $4.94(0.103)$ & $1.04\cdot10^4(1.84\cdot10^2)$ \\
$\alpha=90^{\circ}$ & $5.05(0.114)$ & $1.04\cdot10^4(1.73\cdot10^2)$ \\
\end{tabular}}
\label{suptab: figS3}
\end{table}

\newpage
\subsection{Reconstruction for temporal resolution $\Delta t=2\text{TR}=11$ms}
\begin{figure}[H]
  \centering
    \includegraphics[width=\textwidth]{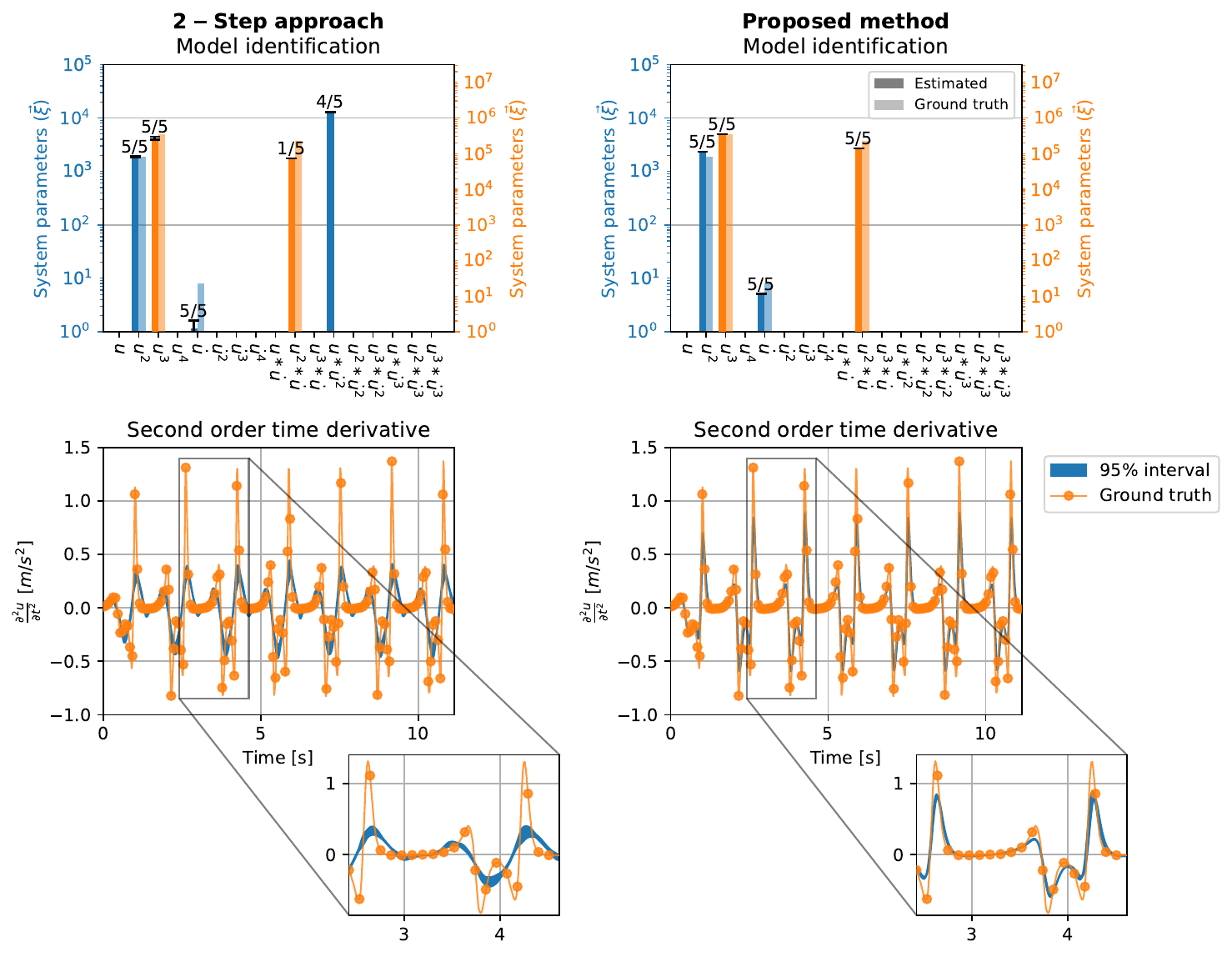}
  \caption{Model identification (displayed as in figure 4 of the paper) and related second order time derivative for dynamic 5, reconstructed using the 2-step approach and the proposed method at an effective temporal resolution of $\Delta t=2\text{TR}=11$ms. The specific values for the reconstructed system parameters are reported in table S6. The color of the bars match the color of the corresponding $y$-axis. The numbers above the bars refer to the amount of times the candidate basis operators are selected out of 5 repetitions of the experiment.}
  \label{supfig: Windowing}
\end{figure}

\begin{table}[H]
  \centering
  \caption{The reconstructed parameters and their respective standard deviations in brackets for selected candidate basis functions using the 2-step approach and the proposed method at an effective temporal resolution of $\Delta t=2\text{TR}=11$ms. These values correspond to the bar chart from figure S5.}
 \scalebox{0.7}{
  \begin{tabular}{c c||l l l l l l}
\multicolumn{2}{c||}{} & $u^2$ & $u^3$ & $\dot{u}$  & $u^2*\dot{u}$ & $u*\dot{u}^2$\\
\hline
\multicolumn{2}{c||}{Ground truth value} & $1.875\cdot10^3$ & $3.472\cdot10^5$ & $-8$ & $2.22\cdot10^5$ & $\times$\\ 
\hline
\multirow{2}{*}{2-step approach} & mean (std) & $1.87\cdot10^3(44.6)$ & $2.70\cdot10^5(2.32\cdot10^4)$ & $-1.16(0.46)$ & $7.28\cdot10^4(0)$ & $1.29\cdot10^4(75.1)$ \\
\cline{2-2}
 & \# selected out of 5 & $5$ & $5$ & $5$ & $1$ & $4$ \\
\hline
\multirow{2}{*}{Proposed method} & mean (std) & $2.36\cdot10^3(9.67)$ & $3.48\cdot10^5(8.28\cdot10^2)$ & $-5.08(0.044)$ & $1.39\cdot10^5(1.32\cdot10^3)$ & $\times$ \\
\cline{2-2}
 & \# selected out of 5 & $5$ & $5$ & $5$ & $5$ & $0$
\end{tabular}}
\label{suptab: tab5}
\end{table}

\newpage
\setcounter{figure}{0}
\renewcommand{\figurename}{Video}
\subsection{Time-resolved data}
\setlength{\intextsep}{\fill}
\begin{figure}[h!]
  \centering
    \includegraphics[width=\textwidth]{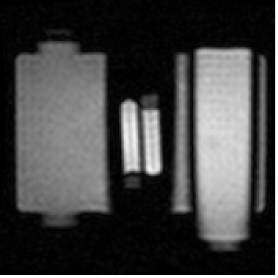}
  \caption{Single frame in the video of the estimates for the time-resolved images for dynamic 5. The time-resolved images are obtained by performing an inverse Fast Fourier Transform (IFFT) on the estimated time-resolved $k$-space data: $\mathcal{F}^{-1}(\boldsymbol{\widehat{m}})$. This is followed by a geometric correction to account for gradient nonlinearities. The full video can be found online.}
  \label{supfig: Video}
\end{figure}